\documentclass[aps,prb,twocolumn]{revtex4-2}
\usepackage{amsmath}
\usepackage{amssymb}
\usepackage{graphicx}% Include figure files
\usepackage{dcolumn}% Align table columns on decimal point
\usepackage{bm}% bold math
\usepackage[colorlinks=true,linkcolor=blue,anchorcolor=blue, citecolor=blue,urlcolor=blue]{hyperref}% add hypertext capabilities
\usepackage[mathlines]{lineno}% Enable numbering of text and display math
\usepackage{ulem}
\newcommand{\eq}[1]{\begin{equation} #1 \end{equation}}
\newcommand{\eqa}[1]{\begin{equation}\begin{aligned} #1 \end{aligned}\end{equation}}
\newcommand{\ua}{\uparrow}
\newcommand{\da}{\downarrow}
\newcommand{\ra}{\rightarrow}
\newcommand{\bfk}{\boldsymbol{k}}
\newcommand{\kb}[1]{\langle #1 \rangle}

\newcommand{\cs}[1]{\left(#1\right)}
\newcommand{\cm}[1]{\left[#1\right] }
\newcommand{\cl}[1]{\left\{#1\right\}}
\newcommand{\bsf}[1]{\boldsymbol{#1}}
\begin{document}
%==============================================================================
\title{Majorana corner modes and tunable patterns in an altermagnet heterostructure}% Force line breaks with \\
\author{Yu-Xuan Li}
\author{Cheng-Cheng Liu}
\email{ccliu@bit.edu.cn}
\affiliation{Centre for Quantum Physics, Key Laboratory of Advanced Optoelectronic Quantum Architecture and Measurement (MOE), School of Physics, Beijing Institute of Technology, Beijing 100081, China}
%\affiliation{ address 3}
%\date{\today}
%=======================Abstract==============================
\begin{abstract}
The mutual competition and synergy of magnetism and superconductivity provide us with a very valuable opportunity to access topological superconductivity and Majorana Fermions. Here, we devise a heterostructure consisting of an $s$-wave superconductor, a 2D topological insulator and an altermagnet, which is classified as the third magnet and featured by zero magnetization but spin polarization in both real and reciprocal spaces. We find that the altermagnet can induce mass terms at the edges that compete with electron pairing, and mass domains are formed at the corners of sample, resulting in zero-energy Majorana corner modes (MCMs). The presence or absence of MCMs can be engineered by only changing the direction of the N\'{e}el vector. Moreover, uniaxial strain can effectively manipulate the patterns of the MCMs, such as moving and interchanging MCMs. Experimental realization, remarkable advantages of our proposal and possible braiding are discussed.
\end{abstract}
\maketitle
%\tableofcontents
%========================Introduction================================================
\textit{\textcolor{black}{Introduction.---}}In the past decade, Majorana zero modes (MZMs) have been extensively researched as building blocks for topological quantum computation~\cite{alicea_new_2012,stanescu_majorana_2013, beenakker_search_2013-1, elliott_colloquium_2015, sarma_majorana_2015, sato_majorana_2016}. Topological superconductors have gained tremendous attention in both experimental and theoretical settings~\cite{qi_topological_2011, alicea_new_2012, sato_topological_2017, lutchyn_majorana_2018, frolov_topological_2020, das_sarma_search_2023} as a promising platform for the realization of MZMs. The key feature that differentiates topological superconductors from trivial ones is the presence of MZMs at the boundaries of the system~\cite{kitaev_unpaired_2001, oreg_helical_2010, lutchyn_majorana_2010} or the core of the vortices~\cite{read_paired_2000, fu_superconducting_2008, sau_generic_2010, alicea_majorana_2010}.

The higher-order topological states of matter have extended our understanding of topological   states~\cite{zhang_surface_2012,sitte_topological_2012,benalcazar_quantized_2017,benalcazar_electric_2017,song__2017,langbehn_reflection-symmetric_2017,ezawa_higher-order_2018,schindler_higher-order_2018,schindler_higher-order_2018-1,khalaf_boundary-obstructed_2021} and provided a new path for the realization of MZMs~\cite{geier_second-order_2018,hsu_majorana_2018,khalaf_higher-order_2018,liu_majorana_2018,shapourian_topological_2018,wang_high-temperature_2018,yan_majorana_2018,zhu_tunable_2018,zhu_second-order_2019,yan_higher-order_2019,zhang_helical_2019,pan_lattice-symmetry-assisted_2019,peng_floquet_2019,zhang_higher-order_2019,gray_evidence_2019,kheirkhah_first-_2020,vu_time-reversal-invariant_2020,schindler_pairing_2020,wu_-plane_2020,ikegaya_tunable_2021,qin_topological_2022,wu_high-_2022}. Compared with first-order conventional topological superconductors, higher-order topological superconductors (HOTSCs) have a special ``bulk-edge" correspondence. Specifically, the codimension ($d_c$) of edge states or end states in conventional topological superconductors is 1, while the $d_c$ of hinge states or corner states in HOTSCs is equal to or greater than 2. Previous studies have proposed several heterojunction systems to realize HOTSCs with MZMs, such as utilizing topological $p$-wave superconductors~\cite{zhu_tunable_2018}, unconventional or conventional superconductor/topological insulator heterojunctions~\cite{yan_majorana_2018,wang_high-temperature_2018,wu_-plane_2020} and ferromagnetic heterojunctions~\cite{liu_majorana_2018,zhang_higher-order_2019}. Although progress has been made in the study of HOTSCs, platforms for tunable MZMs are still difficult to achieve.

\begin{figure}[h]
    \centering
    \includegraphics[width=0.45\textwidth]{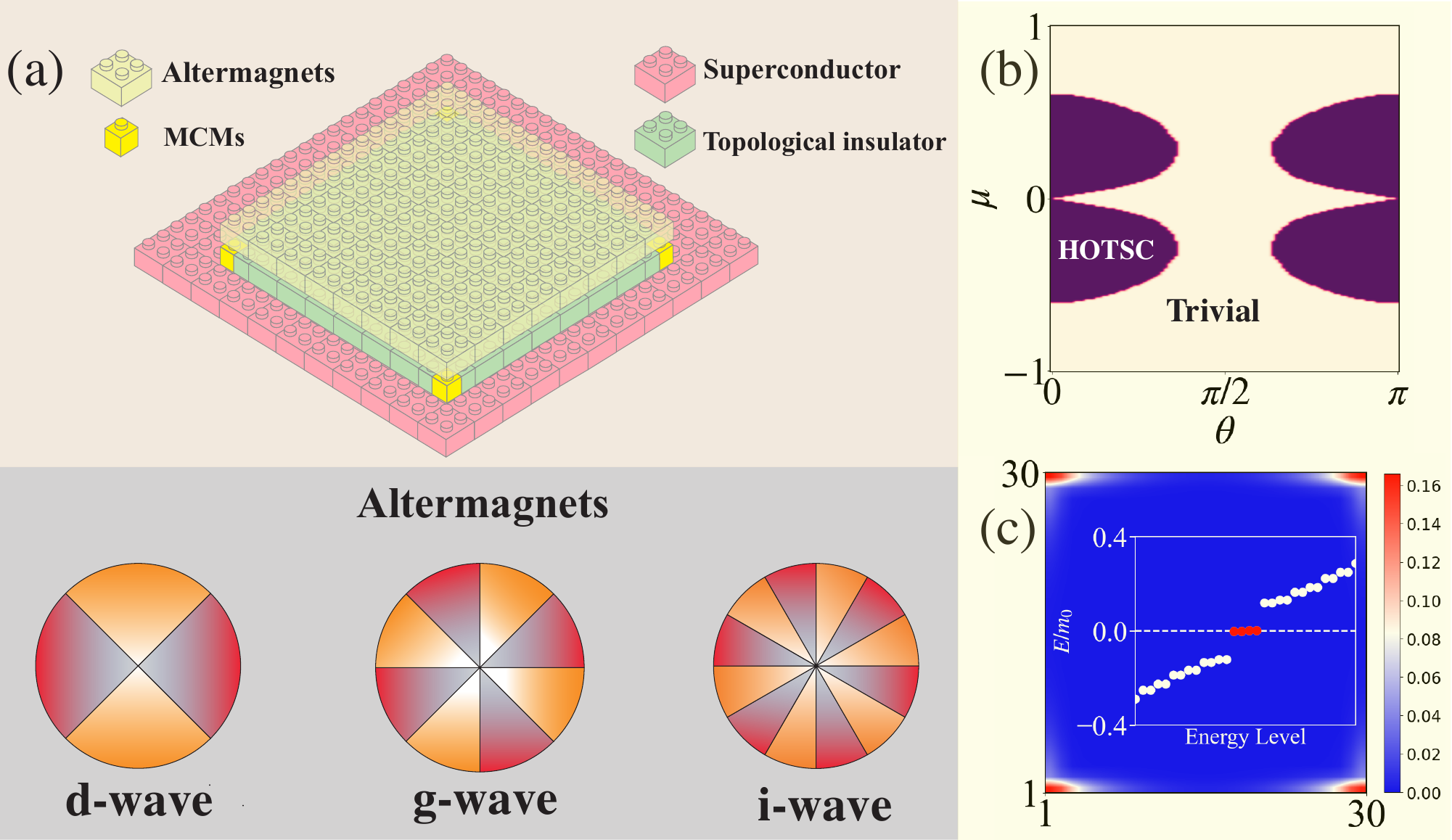}
    \caption{(a) Schematic of the proposed altermanget heterostructure. A two-dimensional topological insulator is sandwiched between an $s$-wave superconductor and an altermagnets. The emerging MZMs are localized at the corners. Altermagnets have various spin-momentum locking with the even-parity wave form, such as $d$-wave, $g$-wave, and $i$-wave. (b) The phase diagram in the plane of N\'{e}el vector polar angle $\theta$ and chemical potential $\mu$.  (c) The real-space wave function distribution with one MZM localized at each corner. (Inset) Plot of the several eigenvalues near  zero energy. The common parameters are $m_0=1.0, t_x=A_x=A_y=1.0, t_y=0.5,\Delta_0=0.5$, and $J_0=1.0$.}\label{fig:ill}
\end{figure}

Recently, a new magnetic phase dubbed altermagnets~\cite{xie_flavor_2023,smejkal_anomalous_2022} was discovered in  several materials, such as  RuO$_2$~\cite{ahn_antiferromagnetism_2019,smejkal_beyond_2022-1}, Mn$_5$Si$_3$~\cite{reichlova_macroscopic_2021} and  KRu$_4$O$_8$~\cite{smejkal_beyond_2022-1}, which breaks the time-reversal symmetry (TRS) despite having zero net magnetization. Interestingly, altermangets with non relativistic highly anisotropic spin splitting~\cite{smejkal_beyond_2022-1} are used to explore some interesting properties~\cite{ma_multifunctional_2021,mazin_prediction_2021,feng_observation_2022,xie_flavor_2023,gonzalez_betancourt_spontaneous_2023,mazin_altermagnetism_2023,sun_andreev_2023,zhang_finite-momentum_2023,zhou_crystal_2023}, such as finite-momentum Cooper pairing, Andreev reflection, Hall effect, etc.

A natural question is whether we can use the highly anisotropic spin splitting and zero magnetization unique to altermangets to design and optimize topological superconductivity, especially higher-order topological superconductivity and the hallmark Majorana modes.

In this work, we investigate the combination of altermagnets with superconductor/topological insulator heterojunctions, as illustrated in Fig. \ref{fig:ill}(a) and give an affirmative answer to the above question. There are several forms of spin splitting in altermagnets with different symmetries, such as $d$-wave, $g$-wave, and $i$-wave~\cite{smejkal_beyond_2022-1}. Without loss of generality, we consider an altermagnets with $d$-wave magnetism as an example, and the other symmetries have a similar analysis procedure~\cite{supp}. Our findings demonstrate that altermagnetism can drive the system into a HOTSC phase, resulting in the emergence of MZMs at the corners of the system, i.e., Majorana corner modes (MCMs).
By adjusting the chemical potential or varying the direction of the N\'{e}el vector, which are both readily achieved experimentally, we can create or annihilate a pair of MCMs. Furthermore, combined with uniaxial stress, we can move and switch MCMs arbitrarily, offering flexible adjustability for further research.
%=================================================================================

\textit{\textcolor{black}{Model.---}} We begin with a general form of the superconducting mean-field Hamiltonian $\mathcal{H}=\frac{1}{2}\sum_{\bfk}\Psi^\dagger_{\bfk}H^{\rm BdG}(\bfk)\Psi_{\bfk}$, where the Nambu spinor basis $\Psi^\dagger_{\bfk}=(c^\dagger_{\bfk\alpha},c_{-\bfk\alpha})$ with $c^\dagger_{\bfk\alpha}$($c_{\bfk\alpha}$) representing the fermion creation (annihilation) operator for the $\alpha$ degree of freedom. The corresponding Bogoliubov-de Gennes (BdG) Hamiltonian is given by
\eqa{
H^{\rm BdG}(\bfk)&=\cs{
\begin{array}{cc}
H_0(\bfk)-\mu\Gamma_{00}&-i\Gamma_{20}\Delta_0\\
i\Gamma_{20}\Delta_0&\mu\Gamma_{00}-H_0^*(-\bfk)
\end{array}
},\\
H_0(\bfk)&=M(\bfk)\Gamma_{03}+A_x\sin k_x\Gamma_{21}+A_y\sin k_y\Gamma_{11}\\
&+J(\bfk)\bsf{s}\cdot\bsf{n}\otimes\sigma_x.\label{ham1}
}
$\Gamma_{ij}= s_i\otimes\sigma_j$, where $s_j$ and $\sigma_j$ act on the spin $(\uparrow,\downarrow)$ and orbital $(a,b)$ degree of freedom, respectively.
The N\'{e}el vector, represented by a unit vector $\bsf{n}=(\sin\theta\cos\varphi,\sin\theta\sin\varphi,\cos\theta)$, can be rotated experimentally~\cite{smejkal_electric_2017-1}. The first line of the Hamiltonian $H_0(\bfk)$ can describe the conventional topological insulator with a nontrivial $\mathcal{Z}_2$ invariant~\cite{kane_z_2005} with $M(\bfk)=(m_0-t_x\cos k_x-t_y\cos k_y)$, while the second line denotes the $d$-wave altermagnetism with $J(\bfk)=J_0(\cos k_x-\cos k_y)$~\cite{smejkal_beyond_2022-1,smejkal_anomalous_2022}. Throughout this work we use the lattice constants as units, and take all the parameters $m_0,t_{x(y)},A_{x(y)},J_0$, and $\Delta_0$ positive. When we ignore the altermagnetism term in the normal states, the Hamiltonian depicts a two-dimensional topological insulator with time-reversal symmetry (TRS)~\cite{bernevig_quantum_2006}. The TRS operation is given by $\mathcal{T}=is_y\mathcal{K}$, where $\mathcal{K}$ is the complex conjugate. When the``mass-ring" $M(\bfk)=0$ [where the $M(\bfk)$ term changes sign] of the topological insulator contains an odd number of time-reversal-invariant momentums (TRIMs), the system has gapless helical edge states protected by the TRS~\cite{fu_topological_2007}.

\begin{figure}[h]
\centering
\includegraphics[width=0.45\textwidth]{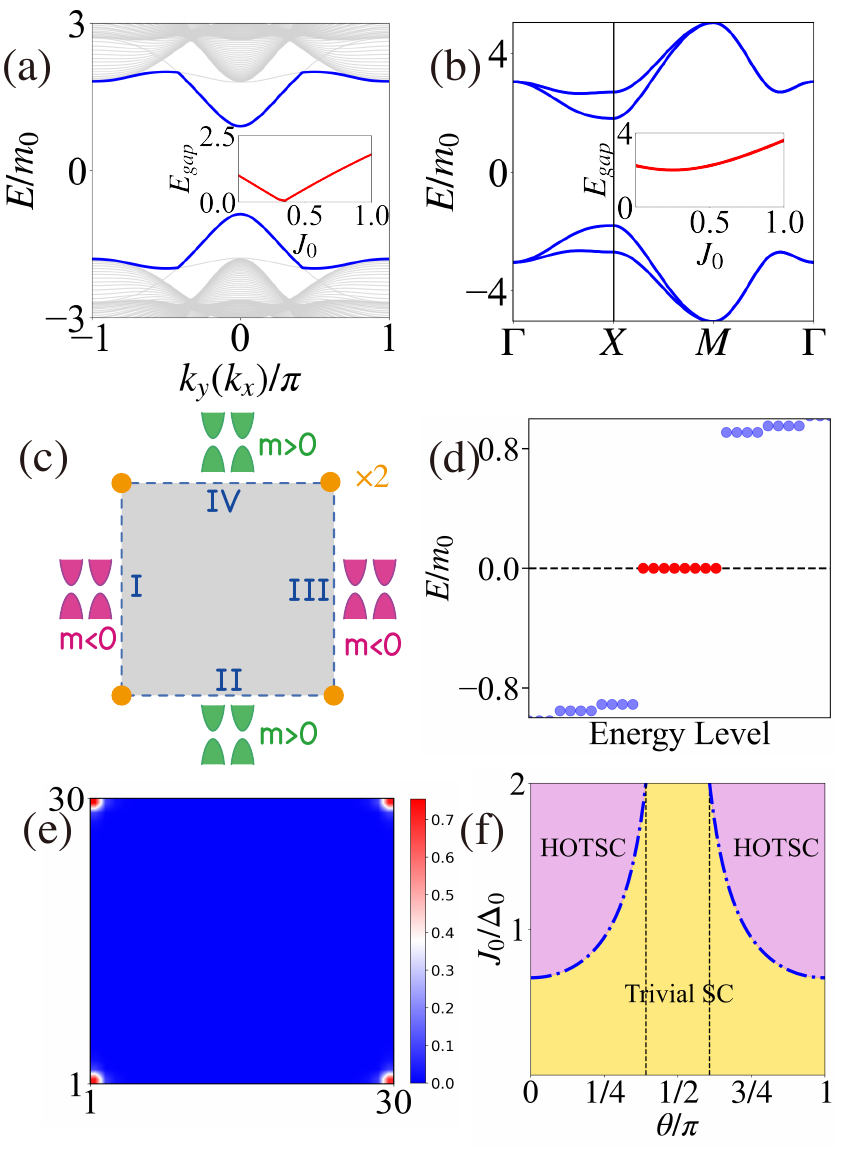}
\caption{(a) Quasiparticle spectrum for a cylindrical geometry. In the presence of altermagnetism and electron pairing, the otherwise helical edge states open a gap. The inset shows the variation of the edge gap with order parameter strength $J_0$ of altermagnets. The closure of the edge states at the critical $J_0=J_0^c$ indicates a topological phase transition at the boundary. (b) Bulk bands along the high symmetry path at $J_0=1.0$. The inset shows the evolution of the bulk gap with $J_0$, and no bulk gap closure is found. (c) Schematic demonstrating the Dirac mass terms of adjacent edges with opposite signs, forming two domain walls.  I, II, III and IV donate four edges. (d) The real-space energy spectrum of an open-boundary square structure of size $L_x=L_y=30$. Eight MZMs emerge (red dots). (e) The distribution of the MZM wavefunction in real space. (f) The phase diagram in the plane of the azimuth angle $\theta$ and $J_0$. The blue dotted lines mark the phase boundary. The common parameters are $m_0=1.0$, $t_x=t_y=A_x=A_y=2.0$, and $\Delta_0=0.5$.}\label{fig:band}
\end{figure}

\textit{\textcolor{black}{Zero Energy MCMs.---}}When proximitized with an $s$-wave superconductor and an altermagnet, the helical edge states will be gapped by breaking the $U(1)$ charge conservation symmetry and TRS, respectively.
 The interplay of the $s$-wave pairing and altermagnet order parameter will drive the system into the HOTSC phase.  We consider the general case and choose the chemical potential and N\'{e}el vector as variable parameters, finding that both have significant effects on the MCMs, as illustrated in the phase diagram [Fig. \ref{fig:ill}(b)]. One can find that there are no MCMs when the N\'{e}el vector is close to the in-plane direction. On the contrary, when the N\'{e}el vector has a certain out-of-plane component, it favors the formation of MCMs. In the  HOTSC phase, there are one MCM at each corner, as shown in Fig. \ref{fig:ill}(c).

To get an intuitive picture of the MCMs, we fix the chemical potential $\mu=0$ and N\'{e}el vector along the $z$ direction ($\theta=0$) for simplicity. When $\Delta_0\neq0, J_0=0$, the TRS of the system still holds, but the broken $U(1)$ symmetry makes the helical edge states open a gap, and the system is a trivial superconductor~\cite{wu_-plane_2020}. With the altermagnetism turned on and the increase of   $J_0$, the edge-state gap along the direction of $x$($y$) will be closed and reopened again, as shown in Fig. \ref{fig:band}(a). However, during the process, the bulk state is always fully gapped [Fig. \ref{fig:band}(b)]. These indicates a possible higher-order topological phase transition ~\cite{khalaf_boundary-obstructed_2021}, which can be made more clearly by the edge theory in the next section. After the phase transition, eight MCMs appear with two MCMs at each corner, which are characteristic of the HOTSC phase, as shown in Figs. \ref{fig:band}(d) and  \ref{fig:band}(e). We will see later that the doubling of MCMs here is due to the additional symmetry brought about by the zero chemical potential. In addition, adjusting the N\'{e}el vector will affect the location of higher-order topological phase transitions. With the chemical potential $\mu=0$, we analytically obtained the phase diagram in the plane of the amplitude $J_0$ and the polar angle $\theta$ of the altermagnetic order parameter, as plotted in Fig. \ref{fig:band}(f).
%=================================================================================

\textit{\textcolor{black}{Edge theory.---}}The appearance of MCMs after the higher-order topological phase transition can be intuitively understood by edge theory.  For simplicity, we first consider the case of the chemical potential $\mu=0$ and N\'{e}el vector along the $z$ axis~\cite{supp2}. We expand the Hamiltonian $H^{\rm BdG}(\bfk)$ in Eq.~\eqref{ham1} around  $\bsf{\Gamma}=(0,0)$ to the second order,
\eqa{
    H^{\rm BdG}(\bfk)&=(m+\frac{t_x}{2}k_x^2+\frac{t_y}{2}k_y^2)\Lambda_{303}+A_xk_xs_y\Lambda_{321}\\
    &+A_yk_y\Lambda_{011}-\frac{J_0}{2}(k_x^2-k_y^2)\Lambda_{331}+\Delta_0\Lambda_{220},
}
with $\Lambda_{ijk}=\tau_i\otimes\Gamma_{jk}$, and $\tau_i$ are Pauli matrices in the particle-hole space. The parameter $m=m_0-t_x-t_y<0$ ensures that the mass ring contains the odd number of TRIMs~\cite{fu_topological_2007}. To distinguish the boundaries, we label the four boundaries using I, II, III, and IV, respectively, as shown in Fig. \ref{fig:band}(c). We first focus on edge I. The momentum $k_x$ is replaced by $-i\partial_x$, and  the Hamiltonian can be decomposed into two parts $H=H_0+H^{\prime}$, which are denoted as $H_0(-i\partial_x,k_y)=(m-t_x\partial_x^2/2)\Lambda_{303}-iA_x\partial_x\Lambda_{321}$ and $H^{\prime}(-i\partial_x,k_y)=A_yk_y\Lambda_{011}+J_0\partial_x^2/2\Lambda_{331}+\Delta_0\Lambda_{220}$, respectively. We solve the eigenequation $H_0\psi_{\alpha}(x)=E_\alpha\psi_{\alpha}(x)$, satisfying the boundary condition $\psi_{\alpha}(0)=\psi_\alpha(+\infty)=0$, and obtain four $E_\alpha=0$ solutions~\cite{supp}. The perturbation part $H^{\prime}$ is projected onto the bases $\psi_{\alpha}$ and reads as $H_{\rm I}=A_yk_y\eta_z-M_{\rm I} \eta_y\gamma_y+J_{\rm I} \eta_x\gamma_z$, where the Pauli matrices $\gamma_i,\eta_j$ are defined in the four zero-energy-states $\psi_\alpha$ space. The mass terms $J_{\rm I}=J_0m/t_x$ and $M_{\rm I}=\Delta_0$ originate from the altermagnetism and electron pairing, respectively~\cite{supp}.

The effective Hamiltonian on the other edges (II-IV) can be obtained similarly~\cite{supp}. For the convenience of the discussion, we define the edge coordinates $l$ counterclockwise, and the effective Hamiltonian can be compactly written as
\eq{
    H^{\rm edge}(l)=-iA(l)\eta_z\partial_l+M(l)\eta_y\gamma_y+J(l)\eta_x\gamma_z\label{edge},
}
where  for the edge coordinate $l=\cl{\text{I-IV}}$, $A(l)=\cl{A_y,A_x,A_y,A_x}, M(l)=\cl{-\Delta_0,\Delta_0,\Delta_0,-\Delta_0}$, and $J(l)=\cl{J_0m/t_x,-J_0m/t_y,J_0m/t_x,-J_0m/t_y}$, respectively.  From Eq.~\eqref{edge} one can obtain that the Dirac masses originated from $s$-wave pairing and the altermagnets are commutable, indicating that they compete with each other on the edges.
The boundary energy spectrum of Eq.~\eqref{edge} reads as $E^{\rm edge}_{\rm I,III}(k_y)=\pm\sqrt{(A_yk_y)^2+(\mp\Delta_0\pm J_{\rm I(III)})^2}$, $E^{\rm edge}_{\rm II,IV}(k_x)=\pm\sqrt{(A_xk_x)^2+(\pm\Delta_0\pm J_{\rm II(IV)})^2}$.
One can find that as $J_0$ increases, the gap of the edge state decreases, and at the critical value $J_0^cm/t_{x(y)}=\pm\Delta_0$, the gap closes that corresponds to the higher-order topological phase transition and is consistent with the numerical results of Fig. \ref{fig:band}.

Furthermore, we can decompose the Hamiltonian \eqref{edge}  as $H=H^+\oplus H^-$ with the help of a conserved quantity $\Pi=\eta_z\gamma_x$. The effective Hamiltonian for two subspaces with $\Pi=\pm1$ can be obtained as
\eqa{
H^+&=iA(l)\tilde{\eta}_z\partial_l+\tilde{M}^+(l)\tilde{\eta}_x,\\
H^-&=iA(l)\tilde{\eta}_z\partial_l+\tilde{M}^-(l)\tilde{\eta}_x,\label{edge2}
}
with Pauli matrices $\tilde{\eta}$ acting in the subspace and $\tilde{M}^\pm(l=\text{I-IV})=\{\pm\Delta_0+J_0|m|/t_x, \mp\Delta_0-J_0|m|/t_y, \mp\Delta_0+J_0|m|/t_x, \pm\Delta_0-J_0|m|/t_y\}$. It can be found that once $J_0$ exceeds the critical value $J_c=\Delta_0 t_0/|m|$ (considering the isotropic case
 $t_x=t_y=t_0$), the signs of the mass terms at the adjacent boundaries in $H^\pm$ are opposite and two mass domain walls are formed at each corner. The zero-energy MCMs formed by $H^+$ and $H^-$ are decoupled from each other, so that two MCMs exist at each corner, as illustrated in Figs.~\ref{fig:band}(c)--\ref{fig:band}(e).
%===========================================================
    \begin{figure}
    \centering
    \includegraphics[width=0.45\textwidth]{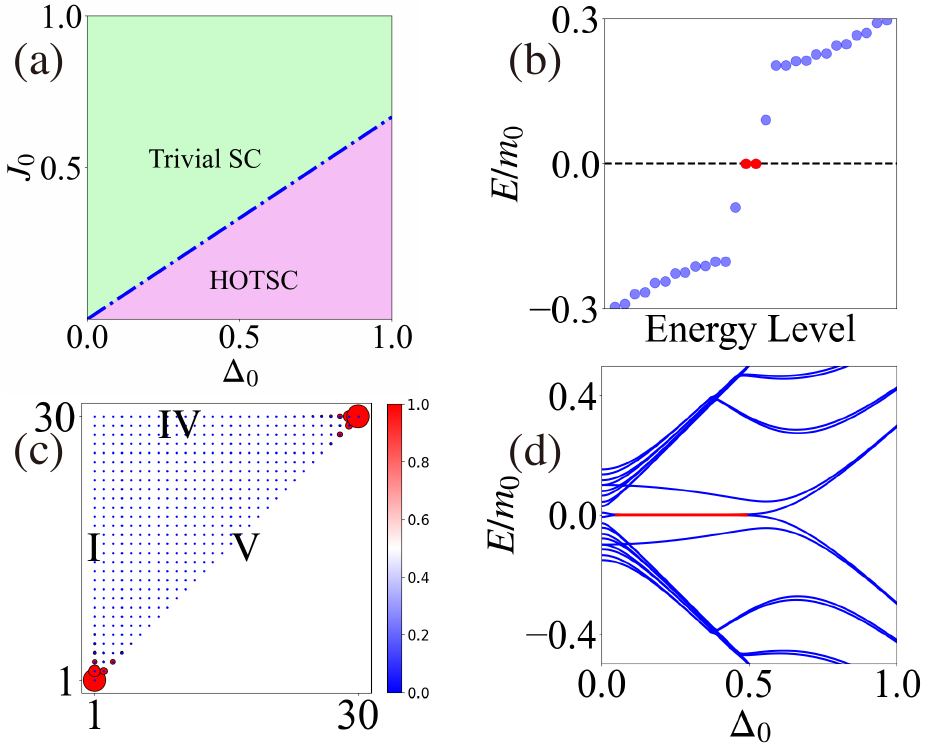}
    \caption{(a) The phase diagram for the isosceles-right-triangle geometry. (b) The energy spectrum of an isosceles triangle structure with two MCMs (red dots) and the density plot of  the two MCMs (c) with V representing the hypotenuse. $\Delta_0$=0.2, $J_0$=0.5. (d) The real-space energy spectrum evolves with electron-pairing amplitude in the isosceles-right-triangle structure with MCMs marked in red existing in a large range. The common parameters are $m_0=1.0$, $t_y=A_y=t_x=A_x=2.0$ and $\mu=0.1$.}\label{fig:geo}
\end{figure}

\textit{\textcolor{black}{Effect of chemical potential.---}}In contrast to the TRS-protected Majorana Kramers pairs~\cite{wang_high-temperature_2018}, only one MZM can stably exist at a given corner without TRS. The chemical potential will give to the edge Hamiltonian~\eqref{edge} a new term $\mu\gamma_z$, which couples the $H^+$ and $H^-$ subblocks. This coupling causes two MCMs from the two sub-blocks at the same corner to mix and disappear. However, two such MCMs are intact if they are located at different corners in spite of nonzero chemical potential. To demonstrate this effect, we start with an isosceles triangle geometry. We first analyze the situation where $t_x=t_y\equiv t_0$, and we will analyze the situation where $t_x\ne t_y$ later. The altermagnetism causes the Dirac mass $J(\alpha)=J_0m\cos(2\alpha)/t_0$ at an edge, where $\alpha$ is the angle between the edge and the edge I. For an edge with $\alpha=\pi/4$, i.e., edge V in Fig. \ref{fig:geo}(c), $J(\alpha)=0$. Similarly we can also get the Dirac mass at edge V, $\tilde{M}_{\rm V}^\pm=\pm\Delta_0$~\cite{supp}.  By carefully analyzing the signs of the mass terms on the three sides of the sample, we find that when $\Delta_0<J_0|m|/t_0$, there is a pair of robust MCMs at two vertices of the triangle sample. In the two subblocks, the signs of the mass terms read $\text{Sign}[\tilde{M}^{+}(\text{I, V, IV})]=\{+,+,-\}$ and  $\text{Sign}[\tilde{M}^{-}(\text{I, V, IV})]=\{+,-,-\}$. Note that two adjacent edges with opposite signs will form a domain wall, so there will be a MCM at the intersection. It is useful to define the corner position, corner i-j, which is defined as the intersection of adjacent edge i and edge j. It can be seen that for $H^+$ subblock, there are two MCMs at corner IV-V and corner IV-I, while for $H^-$ subblock, there are two MCMs at corner I-V and corner I-IV. The nonzero chemical potential will annihilate a pair of MCMs on the same corner (corner IV-I) with only a pair of stable MCMs at two corners (corner V-I, corner V-IV). When $\Delta_0>J_0|m|/t_0$, there is no MCM. These theoretical analyses are also verified by our direct numerical calculations, as shown in Figs.~\ref{fig:geo}(b)--\ref{fig:geo}(d).  We also give the phase diagram in Fig. \ref{fig:geo}(a).
%========================================================================

\textit{\textcolor{black}{Tunable patterns of MCMs.---}}We discover that uniaxial strain can interestingly manipulate the patterns of the MCMs. We first apply a uniaxial stress to a square structure, say along the $x$-direction such that $t_x> t_y$. We find that as the pairing strength increases, the gap of the edge states along the $k_y$ direction of the system is closed and then reopened, but it is always open along the $k_x$ direction, indicating a higher-order topological phase transition~\cite{supp}. The numerical results in Figs.~\ref{fig:chem} (a) and \ref{fig:chem}(b) confirm the higher-order topological phase with four highly localized MZMs at each corner. Such results can be understood through edge theory~\cite{supp}. When $J_0|m|/t_x<\Delta_0<J_0|m|/t_y$, we obtain, for the two subblocks, the signs of the mass terms $\text{Sign}[\tilde{M}^+(l={\rm I-IV})]=\cl{+,-,-,-}$ and
$\text{Sign}[\tilde{M}^-(l={\rm I-IV})]=\cl{-,-,+,-}$, respectively. For $H^+$ sub-block, there are two MCMs at corner I-IV and corner I-II, while for $H^-$ sub-block, there are two MCMs at corner III-II and corner III-IV. Since the four MCMs are located at four different corners,  the nonzero chemical potential does not work, and they can exist stably, as shown in Fig.~\ref{fig:chem} (c). Such an analysis can clearly explain the above numerical results.

We can also apply a uniaxial stress along another direction, such as the $y$ direction, such that $t_x< t_y$. We find a higher-order topological phase transition occurs again with one highly localized MZM at each corner~\cite{supp}. When $J_0|m|/t_y<\Delta_0<J_0|m|/t_x$, similarly, for the two subblocks, the signs of the mass terms read $\text{Sign}[\tilde{M}^{\pm}(l={\rm I-IV})]=\cl{+,\mp,+,\pm}$. There are two MCMs at corner II-I (IV-III) and corner II-III (IV-I) for $H^+$ ($H^-$) subblock. Since the four MCMs are located at four different corners, they can also exist stably despite the non-zero chemical potential, as shown in Fig.~\ref{fig:chem} (d).

It is worth noting that comparing Figs.~\ref{fig:chem}(c) and \ref{fig:chem}(d), one can find that by changing the uniaxial stress direction, we can interchange two MCMs at corner IV-I and corner II-III. Furthermore, we can also freely move a pair of MCMs spatially to locate at any two vertices of the isosceles structure by applying uniaxial stress~\cite{supp}. The intriguing properties with tunable patterns of MCMs hold promise for the implementation of Majorana braiding~\cite{sarma_majorana_2015} and are particularly useful in the design and fabrication of Majorana systems with desired properties~\cite{alicea_non-abelian_2011}.

\begin{figure}[h]
\centering
\includegraphics[width=0.45\textwidth]{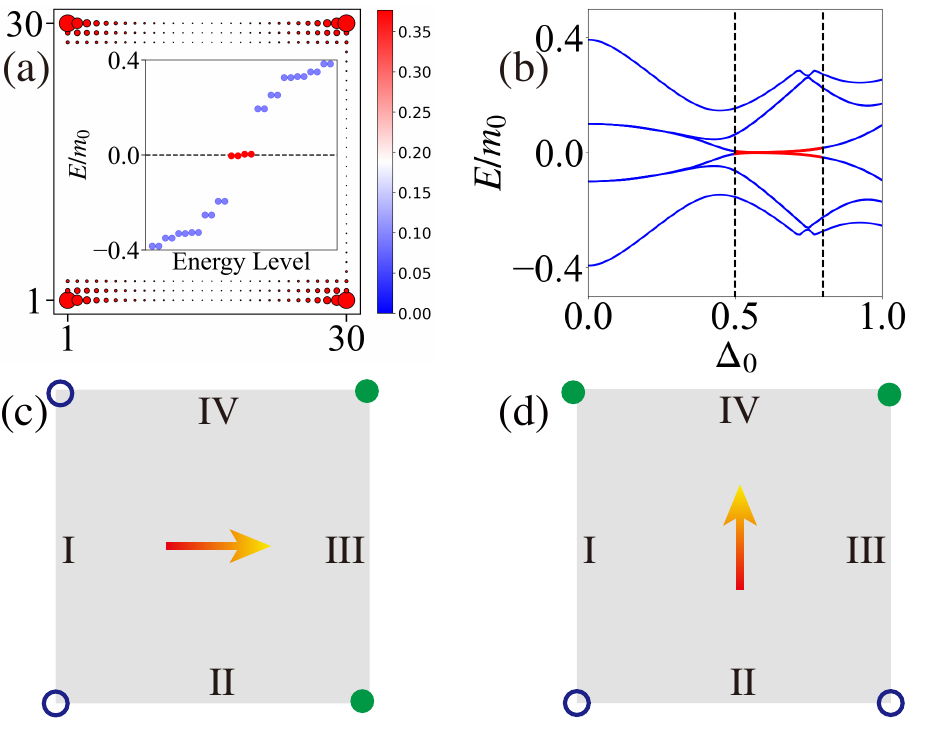}
\caption{Results with nonzero chemical potential $\mu$. (a) The real-space wave function distribution with one MZM localized at each corner. Inset plots the several eigenvalues near the zero energy. (b) The real-space energy spectrum evolves with electron-pairing amplitude in a square structure with MCMs marked in red existing in a range. (c), (d) Tunable MCM patterns. The blue hollow (green solid) circles represent the MCMs from the $H^+$ ($H^-$) block. The arrows indicate the uniaxial stress direction. The common parameters are $m_0=1.0$, $t_y=A_y=1.0$, $t_x=A_x=2.0, J_0=\Delta_0=0.5$, and $\mu=0.1$.}\label{fig:chem}
\end{figure}

%=================================================================================
\textit{\textcolor{black}{Discussion.---}}Although altermagnets have been proposed as the third magnetic phase besides ferromagnets and antiferromagnets for a short time, there are already abundant altermagnets available, such as RuO$_2$~\cite{smejkal_beyond_2022-1,ahn_antiferromagnetism_2019},  KRu$_4$O$_8$~\cite{smejkal_beyond_2022-1} and Mn$_5$Si$_3$~\cite{reichlova_macroscopic_2021}, etc. Recently, the spin splitting in altermagnets has been observed by spin- and angle-resolved photoemission spectroscopy measurement~\cite{zhu_observation_2023}. On the other hand, superconducting/topological insulator heterojunctions have been experimentally realized, and induced electron pairing by proximity effect has been observed in topological insulators~\cite{wang_fully_2013,xu_artificial_2014,zhao_superconducting_2018,shimamura_ultrathin_2018,lupke_proximity-induced_2020}.  Consequently, it is evident that the required material ingredients for implementing the proposal are all experimentally accessible. In the experiment, scanning tunneling microscope can be used to detect MZMs~\cite{STM}, where the differential conductivity peaks at zero energy due to the resonant Andreev reflection are quantized at $2e^2/h$~\cite{law_majorana_2009,wimmer_quantum_2011}. In addition, the fractional Josephson effect~\cite{kitaev_unpaired_2001} is also strong evidence for the existence of a zero-energy Majorana mode, and anomalous response to radio frequency irradiation produced by a $4\pi$-Josephson current could be observed  experimentally~\cite{rokhinson_fractional_2012,wiedenmann_4-periodic_2016}. The existence of MZMs can be proved by combining the above experimental means.

Our proposal has certain remarkable advantages. First, it does not necessitate the use of unconventional superconductors with small electron coherence lengths and an external magnetic field which make experimental realization and confirmation of the existence of MZMs more difficult. Second, by adjusting the direction of the N\'{e}el order parameter of altermagnets, which is readily experimentally, one can create MCMs. In experiments, current, voltage~\cite{godinho_electrically_2018,mahmood_voltage_2021} and spin-orbit torques~\cite{zhang_control_2022} can manipulate and detect the direction of the N\'{e}el vector in antiferromagnetic materials. Last but not least, altermagnets with zero net magnetization have a non-negligible advantage over conventional magnetic materials when combined with superconductivity~\cite{blamire_interface_2014}. All of these make experimental realization of the proposal very possible and opening a new path for the realization of MCMs.

In addition, the patterns of MCMs implemented by our scheme are flexible and adjustable. For example, for an isosceles triangular structure, a pair of MCMs can be freely moved to any two vertices of the triangle by adjusting uniaxial strain, and for a square structure, two MCMs can be interchanged by adjusting uniaxial strain. These operations are important for braiding MCMs. In particular, we noticed that the triangular structure hosts a pair of MCMs, so it is equivalent to 1D topological superconductivity nanowires~\cite{oreg_helical_2010, lutchyn_majorana_2010}, and thus can be used to construct $T$ junctions~\cite{alicea_non-abelian_2011} or $Y$ junctions~\cite{Clarke_prb_2011,Hyart_prb_2013,Wu2014_braiding} for braiding MCMs to exhibit the non-Abelian statistics and potential for topological quantum computation.

%We would like to thank *** for helpful discussions.
The work is supported by the National Key R $\&$ D Program of China (Grant No. 2020YFA0308800) and NSF of China (Grant No. 11922401).

%------------------------------------------------------------------------
%\bibliography{ref}
%==========================================================================================
%merlin.mbs apsrev4-1.bst 2010-07-25 4.21a (PWD, AO, DPC) hacked
%Control: key (0)
%Control: author (72) initials jnrlst
%Control: editor formatted (1) identically to author
%Control: production of article title (-1) disabled
%Control: page (0) single
%Control: year (1) truncated
%Control: production of eprint (0) enabled
%
%===========================================================================================
\clearpage
\onecolumngrid
\begin{center}
    \textbf{\large Supplementary material for ``Majorana corner modes and tunable patterns in an altermagnet heterostructure''}\\[.2cm]
    Yu-Xuan Li and  Cheng-Cheng Liu \\[.1cm]
    {\itshape Centre for Quantum Physics, Key Laboratory of Advanced Optoelectronic Quantum Architecture and Measurement (MOE), School of Physics, Beijing Institute of Technology, Beijing 100081, China}
\end{center}
%---------------------------------------------------------------------------
\maketitle
\setcounter{equation}{0}
\setcounter{section}{0}
\setcounter{figure}{0}
\setcounter{table}{0}
\setcounter{page}{1}
\renewcommand{\theequation}{S\arabic{equation}}
\renewcommand{\thesection}{ \Roman{section}}

\renewcommand{\thefigure}{S\arabic{figure}}
\renewcommand{\thetable}{\arabic{table}}
\renewcommand{\tablename}{Supplementary Table}

\renewcommand{\bibnumfmt}[1]{[S#1]}
\renewcommand{\citenumfont}[1]{#1}
\makeatletter

\maketitle

\setcounter{equation}{0}
\setcounter{section}{0}
\setcounter{figure}{0}
\setcounter{table}{0}
\setcounter{page}{1}
\renewcommand{\theequation}{S-\arabic{equation}}
\renewcommand{\thesection}{ \Roman{section}}

\renewcommand{\thefigure}{S\arabic{figure}}
\renewcommand{\thetable}{\arabic{table}}
\renewcommand{\tablename}{Supplementary Table}

\renewcommand{\bibnumfmt}[1]{[S#1]}
\makeatletter

\maketitle

%----------------------------------------------------------------------------
The supplementary material contains three parts.  The derivation of the edge Hamiltonian of the square structure when the Ne\'{e}l vector is along any direction and the phase diagram are given in Section~\ref{sec:neel}. The edge Hamiltonian of the isosceles structure, when the Ne\'{e}l vector is along the $z$-direction, is given in Section~\ref{sec:neel}. The effective boundary Hamiltonian for the heterostructures of altermagnets with $g$-wave and $i$-wave spin splitting is derived in Section~\ref{sec:neel}.  In Section~\ref{sec:chem},  the effect of anisotropy and chemical potential on the MCMs in the square structure is studied. In Section~\ref{sec:braiding}, we manipulate the pattern of MCMs by uniaxial stresses. We can interchange MCMs in a square structure and move the pair of MCMs spatially to locate at any two vertices of an isosceles structure.

\section{edge theory of Neel vector field along arbitrary direction}\label{sec:neel}
In this section, we analyze both the general Ne\'{e}l vector $\bsf{n}$ and $d$-wave-like altermagnetism, as previously explored in  Refs.~\cite{smejkal_beyond_2022-1,smejkal_anomalous_2022}. The associated bulk Hamiltonian can be expressed as
\eqa{
    H^{\rm BdG}(\bfk)&=\cs{
        \begin{array}{cc}
            H_0(\bfk)-\mu_0&-is_y\Delta_0\\
            is_y\Delta_0&\mu_0-H_0^*(-\bfk)
        \end{array}
    },\\
    H_0(\bfk)&=(m_0-t_x\cos k_x-t_y\cos k_y)\sigma_z+A_x\sin k_xs_y\sigma_x+A_y\sin k_ys_x\sigma_x+\bsf{J}(\bfk)\bsf{s}\cdot\bsf{n}\sigma_x,\\
    J_x(\bfk)&=J^x_0(\cos k_x-\cos k_y), J_y(\bfk)=J^y_0(\cos k_x-\cos k_y), J_z(\bfk)=J^z_0(\cos k_x-\cos k_y).\label{eq:bdg}
}
The bulk Hamiltonian Eq.~\eqref{eq:bdg} we consider includes Pauli matrices $s_i$ and $\sigma_j$, which act on the spin $(\ua,\da)$ and orbit $(a,b)$ degrees of freedom, respectively. The unit vector $\bsf{n}=(\sin\theta\cos\varphi,\sin\theta\sin\varphi,\cos\theta)$ represents the direction of Ne\'{e}l vector, with $\theta$ and $\varphi$ denoting the polar and azimuthal angles, respectively (Fig.~\ref{fig:rot}(a)). Specifically, the magnitudes of the components in each direction are $J_0^x=J_0\sin\theta\cos\varphi$, $J_0^y=J_0\sin\theta\sin\varphi$, and $J_0^z=J_0\cos\theta$.
\begin{figure}[h]
    \centering
    \includegraphics[scale=0.8]{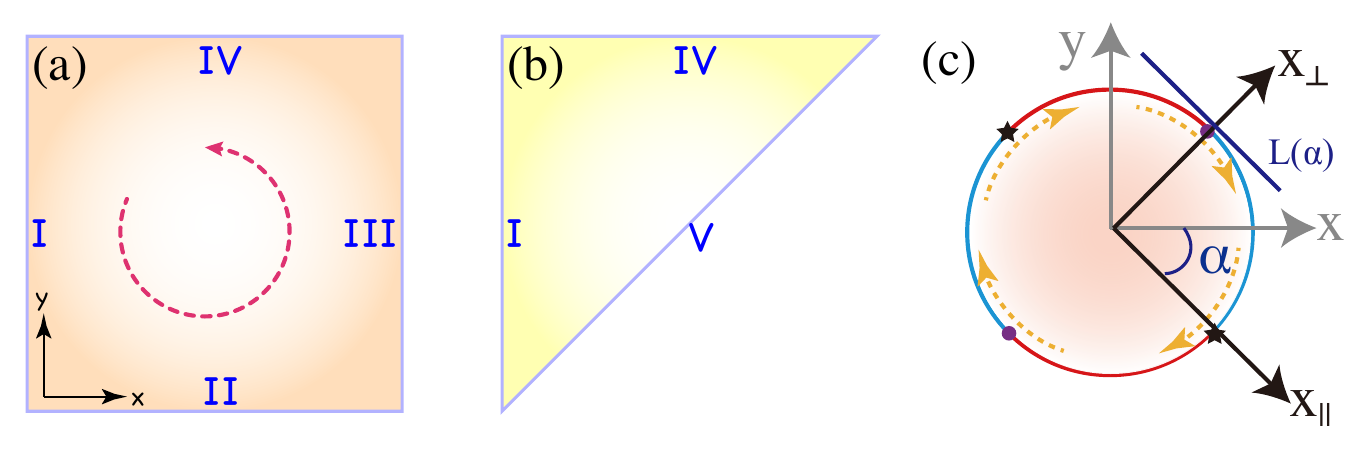}
    \caption{(a) The schematic shows the edge coordinate $l$ with the counterclockwise direction selected as the positive direction. Edges I, II, III, and IV are labeled for use in the edge theory. (b) Schematic of an isosceles triangle structure with the hypotenuse marked as V and the rest of the edges identical to the square structure. (c) Schematic diagram of the rotated coordinate system and the tangential boundary.}\label{sfig:ill}
\end{figure}
%------------------------------------------------------------------------------
\subsection{Square geometry}
In this subsection, we present in detail the edge theory of the square geometry (Fig.~\ref{sfig:ill}(a)).  For simplicity, we set the chemical potential as $\mu=0$ and expand the Hamiltonian in Eq.~(\ref{eq:bdg}) around $\bfk=(0,0)$ to the second order
\eqa{
    H(\bfk)&=(m+\frac{t_x}{2}k_x^2+\frac{t_y}{2}k_y^2)\sigma_z\tau_z+A_xk_xs_y\sigma_x\tau_z+A_yk_ys_x\sigma_x\\
    &-\frac{1}{2}J_0^x(k_x^2-k_y^2)s_x\sigma_x\tau_z-\frac{1}{2}J_0^y(k_x^2-k_y^2)s_y\sigma_x\\
    &-\frac{1}{2}J_0^z(k_x^2-k_y^2)s_z\sigma_x\tau_z+\Delta_0s_y\tau_y.
}
To ensure the topological insulator has helical edge states we set $m=m_0-t_x-t_y<0$. To facilitate discussing different edges we adopt the counterclockwise direction as positive and label the four edges of a square geometry as edge I, edge II, edge III, and edge IV, as shown in Fig.~\ref{fig:ill}(a).

We begin by focusing on the region $x\in [0,+\infty)$ corresponding to edge I. By decomposing the Hamiltonian into $H=H_0+H^\prime$ and applying $k_x\ra -i\partial_x$, we obtain
\eqa{
    H_0(-i\partial_x)&=(m-\frac{t_x}{2}\partial_x^2)\sigma_z\tau_z-iA_x\partial_x\sigma_xs_y\tau_z,\\
    H^\prime(-i\partial_x,k_y)&=A_yk_y\sigma_xs_x+J_x(-i\partial_x)\sigma_xs_x\tau_z+J_y(-i\partial_x)\sigma_xs_y+J_z(-i\partial_x)\sigma_xs_z\tau_z+\Delta_0s_y\tau_y,
}
where we ignore the $k_y^2$ terms. $J_{\alpha,(\alpha=x,y,z)}(-i\partial_x)=J_0^{\alpha,(\alpha=x,y,z)}\partial_x^2/2$. When the pairing and altermagnetism are small compared to the energy gap, $H^\prime$ is regarded as a perturbation. By solving $H_{0}\psi_\alpha(x)=E_\alpha\psi_\alpha(x)$ for $E_\alpha=0$ with the boundary condition $\psi_{\alpha}(0)=\psi_{\alpha}(+\infty)=0$, we obtain four zero-energy eigenstates
\begin{equation}
    \psi_\alpha(x)=\mathcal{N}_x\sin(\kappa_1x)e^{-\kappa_2x}e^{ik_yy}\chi_\alpha,
\end{equation}
with normalization constant  $|\mathcal{N}_x|^2=4|\kappa_2(\kappa_1^2+\kappa_2^2)/\kappa_1^2|$, where $\kappa_1=\sqrt{|2m/t_x|-(A_x^2/t_x^2)}, \kappa_2=A_x/t_x$.

The spinor component of the eigenstate $\chi_\alpha$ satisfies $\tau_0s_y\sigma_y\chi_\alpha=-\chi_\alpha$. Furthermore, we denote the eigenvectors with the eigenvalue equal to $-1$ as $\xi_{\alpha}$, which satisfy $\tau_0s_y\sigma_y\xi_{\alpha}=-\xi_{\alpha}$
\begin{equation}
    \begin{aligned}
        \xi_1&=|\tau_z=+1\rangle\otimes|\sigma_y=+1\rangle\otimes|s_y=-1\rangle,\\
        \xi_2&=|\tau_z=+1\rangle\otimes|\sigma_y=-1\rangle\otimes|s_{y}=+1\rangle,\\
        \xi_3&=|\tau_z=-1\rangle\otimes|\sigma_y=+1\rangle\otimes|s_{y}=-1\rangle,\\
        \xi_4&=|\tau_z=-1\rangle\otimes|\sigma_y=-1\rangle\otimes|s_{y}=+1\rangle.
    \end{aligned}
\end{equation}
We can obtain $\chi_i$ by combining the eigenvectors $\xi_i$
\eqa{
    \chi_1&=(\xi_3+\xi_4)/2=\frac{1}{\sqrt{2}}\cs{0,0,0,0,1,0,0,1}^T,\\
    \chi_2&=-i(\xi_3-\xi_4)/2=\frac{1}{\sqrt{2}}\cs{0,0,0,0,-1,1,0,0}^T,\\
    \chi_3&=(\xi_1+\xi_2)/2=\frac{1}{\sqrt{2}}\cs{1,0,0,1,0,0,0,0}^T,\\
    \chi_4&=-i(\xi_1-\xi_2)/2=\frac{1}{\sqrt{2}}\cs{0,-1,1,0,0,0,0,0}^T.
}
Using these bases, the matrix elements of $H_{p}$ can be expressed as
\begin{equation}
    H_{\text{I},\alpha\beta}=\int_{0}^{\infty}dx\psi_\alpha^*(x)H^\prime(-i\partial_x,k_y)\psi_\beta(x).
\end{equation}
After calculations, we obtain
\eq{
    H_{\rm I}(k_y)=A_yk_y\eta_z-M_{\rm I}\gamma_y\eta_y +J^z_{\rm I}\gamma_z\eta_x-J^x_{\rm I}\gamma_z\eta_z,
}
where $\eta_i$ and $\gamma_j$ are Pauli  matrices in bases of $\psi_\alpha$ and the coefficients are
\begin{equation}
    \begin{aligned}
        J^z_\text{I}&=\frac{J_0^z}{2}\int_{0}^{\infty}dx\psi_\alpha^*(x)(\partial_x^2)\psi_\alpha=\frac{J_0^zm}{t_{x}},\quad J^x_\text{I}=\frac{J_0^x}{2}\int_{0}^{\infty}dx\psi_\alpha^*(x)(\partial_x^2)\psi_\alpha(x)=\frac{J_0^xm}{t_{x}},\\
        M_{\rm I}&=\Delta_0\int_{0}^{\infty}dx\psi_\alpha^*(x)\psi_\alpha(x)=\Delta_0.
    \end{aligned}
\end{equation}
It should be noted that the $y$ component of Ne\'{e}l vector does not impact the behavior of the helical edge states and the  $x$ component of Ne\'{e}l vector does not induce a gap on the helical edge states of the topological insulator.

For edge III, the occupied space is $x\in(-\infty,0]$. Solving the eigenequation $H_0\psi_\alpha(x)=0$ with the boundary condition  $\psi_{\alpha}(0)=\psi_{\alpha}(-\infty)=0$, we obtain the four solutions
\begin{equation}
    \psi_\alpha(x)=\mathcal{N}_x\sin(\kappa_1x)e^{\kappa_2x}e^{ik_yy}\chi_\alpha.
\end{equation}
Unlike edge I, the spinor parts of the wave function  are the eigenstates of $\tau_0s_y\sigma_y$ with the eigenvalue equal to $+1$, and they are denoted as
\begin{equation}
    \begin{aligned}
        \xi_1&=|\tau_z=+1\rangle\otimes|\sigma_y=-1\rangle\otimes|s_y=-1\rangle,\\
        \xi_2&=|\tau_z=+1\rangle\otimes|\sigma_y=+1\rangle\otimes|s_{y}=+1\rangle,\\
        \xi_3&=|\tau_z=-1\rangle\otimes|\sigma_y=-1\rangle\otimes|s_{y}=-1\rangle,\\
        \xi_4&=|\tau_z=-1\rangle\otimes|\sigma_y=+1\rangle\otimes|s_{y}=+1\rangle.
    \end{aligned}
\end{equation}
We combine the eigenvectors $\xi_i$ to obtain $\chi_i$
\begin{equation}
    \begin{aligned}
        \chi_1&=(\xi_3+\xi_4)/2=\frac{1}{\sqrt{2}}\cs{0,0,0,0,-1,0,0,1}^T,\\
        \chi_2&=-i(\xi_3-\xi_4)/2=\frac{1}{\sqrt{2}}\cs{0,0,0,0,1,1,0,0}^T,\\
        \chi_3&=(\xi_1+\xi_2)/2=\frac{1}{\sqrt{2}}\cs{-1,0,0,1,0,0,0,0}^T,\\
        \chi_4&=-i(\xi_1-\xi_2)/2=\frac{1}{\sqrt{2}}\cs{0,1,1,0,0,0,0,0}^T.
    \end{aligned}
\end{equation}
After choosing the basis function as $\psi_\alpha$, the matrix elements of $H^\prime$ under these bases are expressed as
\begin{equation}
    H_{\text{III},\alpha\beta}=\int_{-\infty}^{0}dx\psi_\alpha^*(x)H^\prime(-i\partial_x,k_y)\psi_\beta(x).
\end{equation}
Through algebraic operations, one can obtain the edge Hamiltonian for edge III
\eq{
    H_{\rm III}(k_y)=-A_yk_y\eta_z+M_{\rm III}\gamma_y\eta_y+J^z_{\rm III}\gamma_z\eta_x+J^x_{\rm III}\gamma_z\eta_z,
}
where $\eta_i$ and $\gamma_j$ are Pauli  matrices in bases of $\psi_\alpha$. And coefficients are
\begin{equation}
    \begin{aligned}
        J^z_\text{III}&=\frac{J_0^z}{2}\int^{0}_{-\infty}dx\psi_\alpha^*(x)(\partial_x^2)\psi_\alpha=\frac{J_0^zm}{t_{x}},\quad J^x_\text{III}=\frac{J_0^x}{2}\int^{0}_{-\infty}dx\psi_\alpha^*(x)(\partial_x^2)\psi_\alpha(x)=\frac{J_0^xm}{t_{x}},\\
        M_{\rm III}&=\Delta_0\int^{0}_{-\infty}dx\psi_\alpha^*(x)\psi_\alpha(x)=\Delta_0.
    \end{aligned}
\end{equation}

Next, we consider the edges II and IV. The process is similar to the edge I.  For the edge II, we can decompose the Hamiltonian into two parts
\eqa{
    H_{0}(k_x,-i\partial_y)&=(m-t_y\partial_y^2/2)\tau_{z}\sigma_{z}-iA_ys_{x}\sigma_{x}\partial_y,\\
    H^\prime(k_x,-i\partial_y)&=A_xk_x\tau_{z}s_{y}\sigma_{x}+J_x(-i\partial_y)\sigma_xs_s\tau_z+J_y(-i\partial_y)\sigma_xs_y+J_z(-i\partial_y)\sigma_xs_z\tau_z+\Delta_0s_y\tau_y,
}
where the $k_x^2$ terms are ignored. $J_{\alpha,(\alpha=x,y,z)}(-i\partial_y)-J_0^{\alpha,(\alpha=x,y,z)}\partial_y^2/2$. The space $y\in [0,+\infty)$ is occupied and we solve eigenequation $H_{0}\psi_\alpha(y)=E_\alpha\psi_\alpha(y)$ under the boundary condition $\psi_{\alpha}(0)=\psi_{\alpha}(+\infty)=0$. We obtain the four zero-energy solutions
\begin{equation}
    \psi_\alpha(y)=\mathcal{N}_y\sin(\kappa_3y)e^{-\kappa_4y}e^{ik_xx}\chi_\alpha.
\end{equation}
Here, the normalization parameter is given by $|\mathcal{N}_y|^2=4|\kappa_4(\kappa_3^2+\kappa_4^2)/\kappa_3^2|$, where $\kappa_3=\sqrt{(2|m|/t_y)-(A_y^2/t_y^2)}, \kappa_4=(A_y/t_y)$.
We obtain the spinor parts of the eigenstates $\psi_\alpha$, which satisfy $\tau_{z}s_{x}\sigma_y\chi_\alpha=-\chi_\alpha$.
The eigenstates of $\tau_{z}s_{x}\sigma_y$ with the eigenvalue equal to $-1$ are expressed as
\begin{equation}
    \begin{aligned}
        \xi_1&=|\tau_z=+1\rangle\otimes|\sigma_y=+1\rangle\otimes|s_{x}=-1\rangle,\\
        \xi_2&=|\tau_z=+1\rangle\otimes|\sigma_y=-1\rangle\otimes|s_{x}=+1\rangle,\\
        \xi_3&=|\tau_z=-1\rangle\otimes|\sigma_y=+1\rangle\otimes|s_{x}=+1\rangle,\\
        \xi_4&=|\tau_z=-1\rangle\otimes|\sigma_y=-1\rangle\otimes|s_{x}=-1\rangle.
    \end{aligned}
\end{equation}
We combine the eigenvectors $\xi_i$ to obtain $\chi_i$
\eqa{
    \chi_1&=(\xi_3+\xi_4)/2=\frac{1}{\sqrt{2}}\cs{0,0,0,0,-i,0,0,1}^T,\\
    \chi_2&=(\xi_3-\xi_4)/2=\frac{1}{\sqrt{2}}\cs{0,0,0,0,0,i,1,0}^T,\\
    \chi_3&=(\xi_1+\xi_2)/2=\frac{1}{\sqrt{2}}\cs{i,0,0,1,0,0,0,0}^T,\\
    \chi_4&=(\xi_2-\xi_1)/2=\frac{1}{\sqrt{2}}\cs{0,-i,1,0,0,0,0,0}^T.
}
The perturbation matrix elements read
\begin{equation}
    H_{\text{II},\alpha\beta}=\int_{0}^{+\infty}dy\psi_\alpha^*(y)H^\prime(k_x,-i\partial_y)\psi_\beta(y).\label{per}
\end{equation}
After some algebraic calculations, we obtain
\eq{
    H_{\text{II}}(k_{x})=-A_{x}k_{x}\eta_{z}-M_{\text{II}}\gamma_{x}\eta_{y}+J^z_{\rm II}\gamma_z\eta_x+J^y_{\rm II}\gamma_z\eta_z,
}
where $\eta_i$ and $\tau_j$ are Pauli  matrices in bases of $\psi_\alpha$ and
\begin{equation}
    \begin{aligned}
        J^z_\text{II}&=-\frac{J_0^z}{2}\int_{0}^{\infty}dy\psi_\alpha^*(y)(\partial_y^2)\psi_\alpha(y)=-\frac{J_0^zm}{t_y},\quad J^y_\text{II}=-\frac{J_0^z}{2}\int_{0}^{\infty}dy\psi_\alpha^*(y)(\partial_y^2)\psi_\alpha(y)=-\frac{J_0^ym}{t_y},\\
        M_\text{II}&=\Delta_{0}\int_{0}^{\infty}dy\psi_\alpha^*(y)\psi_\alpha(y)=\Delta_0.
    \end{aligned}
\end{equation}

Finally, for edge IV, we need to solve the eigenvalue equation $H_{0}\psi_\alpha(y)=E_\alpha\psi_\alpha(y)$ under boundary condition $\psi_{\alpha}(0)=\psi_{\alpha}(-\infty)=0$, we obtain the four zero-energy solutions
\begin{equation}
    \psi_\alpha(y)=\mathcal{N}_y\sin(\kappa_3y)e^{\kappa_4y}e^{ik_xx}\chi_\alpha.
\end{equation}
The eigenvectors $\chi_\alpha$ satisfy $\tau_{z}s_{x}\sigma_y\chi_\alpha=\chi_\alpha$. For $\tau_{z}s_{x}\sigma_y$, the eigenvectors with the eigenvalue equal to $+1$ are denoted as
\begin{equation}
    \begin{aligned}
        \xi_1&=|\tau_z=+1\rangle\otimes|\sigma_y=-1\rangle\otimes|s_{x}=-1\rangle,\\
        \xi_2&=|\tau_z=+1\rangle\otimes|\sigma_y=+1\rangle\otimes|s_{x}=+1\rangle,\\
        \xi_3&=|\tau_z=-1\rangle\otimes|\sigma_y=-1\rangle\otimes|s_{x}=+1\rangle,\\
        \xi_4&=|\tau_z=-1\rangle\otimes|\sigma_y=+1\rangle\otimes|s_{x}=-1\rangle.
    \end{aligned}
\end{equation}
We combine the eigenvectors $\xi_i$ to obtain $\chi_i$ with
\begin{equation}
    \begin{aligned}
        \chi_1&=(\xi_3+\xi_4)/2=\frac{1}{\sqrt{2}}\cs{0,0,0,0,i,0,0,1}^T,\\
        \chi_2&=(\xi_3-\xi_4)/2=\frac{1}{\sqrt{2}}\cs{0,0,0,0,0,-i,1,0}^T,\\
        \chi_3&=(\xi_1+\xi_2)/2=\frac{1}{\sqrt{2}}\cs{-i,0,0,1,0,0,0,0}^T,\\
        \chi_4&=(\xi_2-\xi_1)/2=\frac{1}{\sqrt{2}}\cs{0,i,1,0,0,0,0,0}^T,
    \end{aligned}
\end{equation}
and the perturbation matrix elements read
\begin{equation}
    H_{\text{IV},\alpha\beta}=\int_{-\infty}^{0}dy\psi_\alpha^*(y)H^\prime(k_x,-i\partial_y)\psi_\beta(y).
\end{equation}
After some algebraic calculations, we obtain
\eq{
    H_{\text{IV}}(k_{x})=A_{x}k_{x}\eta_{z}+M_{\text{IV}}\gamma_{x}\eta_{y}+J^z_{\rm IV}\gamma_z\eta_x-J^y_{\rm IV}\gamma_z\eta_z,
}
where $\eta_i$ and $\tau_j$ are Pauli  matrices in bases of $\psi_\alpha$ and
\begin{equation}
    \begin{aligned}
        J^z_\text{IV}&=-\frac{J_0^z}{2}\int^{0}_{-\infty}dy\psi_\alpha^*(y)(\partial_y^2)\psi_\alpha(y)=-\frac{J_0^zm}{t_y},\quad J^y_\text{IV}=-\frac{J_0^z}{2}\int^{0}_{-\infty}dy\psi_\alpha^*(y)(\partial_y^2)\psi_\alpha(y)=-\frac{J_0^ym}{t_y},\\
        M_\text{IV}&=\Delta_{0}\int_{0}^{\infty}dy\psi_\alpha^*(y)\psi_\alpha(y)=\Delta_0.
    \end{aligned}
\end{equation}

In summary, the effective Hamiltonian of the four edges are
\eqa{
    &H_{\rm I}(k_y)=A_yk_y\eta_z-M_{\rm I}\gamma_y\eta_y +J^z_{\rm I}\gamma_z\eta_x-J^x_{\rm I}\gamma_z\eta_z,\\
    &H_{\text{II}}(k_{x})=-A_{x}k_{x}\eta_{z}-M_{\text{II}}\gamma_{x}\eta_{y}+J^z_{\rm II}\gamma_z\eta_x-J^y_{\rm II}\gamma_z\eta_z,\\
    &H_{\rm III}(k_y)=-A_yk_y\eta_z+M_{\rm III}\gamma_y\eta_y+J^z_{\rm III}\gamma_z\eta_x+J^x_{\rm III}\gamma_z\eta_z,\\
    &H_{\text{IV}}(k_{x})=A_{x}k_{x}\eta_{z}+M_{\text{IV}}\gamma_{x}\eta_{y}+J^z_{\rm IV}\gamma_z\eta_x+J^y_{\rm IV}\gamma_z\eta_z.\label{pb1}
}
From the equation \eqref{pb1} we obtain that only the $z$ component of  Ne\'{e}l vector contributes a Dirac mass for the helical edge states. This indicates that the $z$ component plays a crucial role in determining the topological properties of the system. Therefore,  we focus on the $z$ component of the Ne\'{e}l vector and examine its effects on the edge states
\eqa{
    &H_{\rm I}(k_y)=A_yk_y\eta_z-M_{\rm I}\gamma_y\eta_y +J^z_{\rm I}\gamma_z\eta_x,\\
    &H_{\text{II}}(k_{x})=-A_{x}k_{x}\eta_{z}-M_{\text{II}}\gamma_{x}\eta_{y}+J^z_{\rm II}\gamma_z\eta_x,\\
    &H_{\rm III}(k_y)=-A_yk_y\eta_z+M_{\rm III}\gamma_y\eta_y+J^z_{\rm III}\gamma_z\eta_x,\\
    &H_{\text{IV}}(k_{x})=A_{x}k_{x}\eta_{z}+M_{\text{IV}}\gamma_{x}\eta_{y}+J^z_{\rm IV}\gamma_z\eta_x.
}
We introduce a unitary transformation
\begin{equation}
    U=e^{i\frac{\pi}{4}\gamma_z}\otimes\mathbf{1}_{2\times 2},
\end{equation}
where $\mathbf{I}_{2\times2}$ is a 2D identity matrix. Under the unitary transformation, the effective Hamiltonian for edge II and edge IV is transformed to
\eqa{
    H_{\rm II}&=-A_xk_x\eta_z+M_{\rm II} \gamma_y\eta_y+J^z_{\rm II} \gamma_z\eta_x,\\
    H_{\rm IV}&=A_xk_x\eta_z-M_{\rm IV} \gamma_y\eta_y+J^z_{\rm IV} \gamma_z\eta_x.
}

The effective Hamiltonian of the four edges read
\eqa{
    H_{\rm I}&=A_yk_y\eta_z-\Delta_0 \gamma_y\eta_y+\frac{J^z_0m}{t_x} \gamma_z\eta_x,\\
    H_{\rm II}&=-A_xk_x\eta_z+\Delta_0 \gamma_y\eta_y-\frac{J^z_0m}{t_y} \gamma_z\eta_x,\\
    H_{\rm III}&=-A_yk_y\eta_z+\Delta_0 \gamma_y\eta_y+\frac{J^z_0m}{t_x} \gamma_z\eta_x,\\
    H_{\rm IV}&=A_xk_x\eta_z-\Delta_0 \gamma_y\eta_y-\frac{J^z_0m}{t_y} \gamma_z\eta_x.\label{eq:edge}
}

Taking the counterclockwise direction as positive, we can take an 'edge coordinate' $l$ as shown in Fig.~\ref{fig:ill}(a), and the effective Hamiltonian can be expressed as
\begin{equation}
    H_{\text{edge},l}=-iA(l)\eta_{z}\partial_l+M(l)\gamma_{y}\eta_{y}+J(l)\gamma_z\eta_{x},
\end{equation}
with the coefficients $A(l)=\{A_y,A_x,A_y,A_x\}$, $M(l)=\cl{-\Delta_0,\Delta_0,\Delta_0,-\Delta_0}$ and $J(l)=\left\{J_0^zm/t_x,-J_0^zm/t_y\right.$,  $\left. J_0^zm/t_x,-J_0^zm/t_y\right\}$ for $l=\{$I, II, III, IV$\}$, respectively.
\begin{figure}[h]
    \centering
    \includegraphics[scale=0.35]{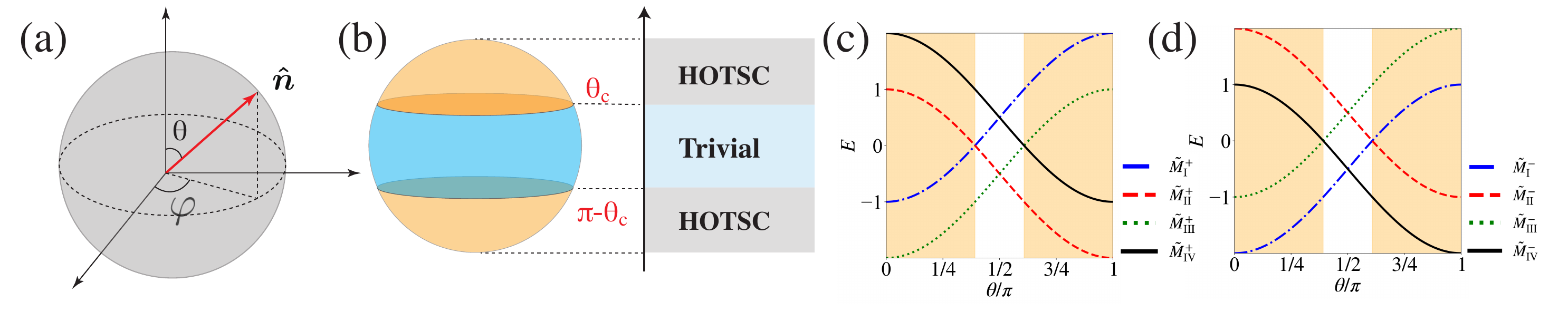}
    \caption{(a) The unit vector $\bsf{\hat{n}}$ of the Ne\'{e}l vector with the polar angle $\theta$ and azimuthal $\varphi$ angles in the spherical coordinates. (b) Phase diagram concerning the polar $\theta$ and azimuthal $\varphi$ angles. Here the critical values $\theta_c$ exist for the polar angle. (c) and (d) The variation of $\tilde{M}^+$ and $\tilde{M}^-$ with the polar angle of the Ne\'{e}l vector, respectively. The Dirac mass terms $\tilde{M}^\pm$ of the adjacent edges in the orange region have opposite signs. Common parameters are: $m_0=1.0, t_x=t_y=A_x=A_y=2.0, \Delta_0=J_0=0.5,\mu=0$.}\label{fig:rot}
\end{figure}

The Hamiltonian $H_{\text{edge},l}$ can be decoupled as $H_{\text{edge},l}=H_{\text{edge},l}^{+}\oplus H_{\text{edge},l}^{-}$ with the help of   $\Pi=\gamma_x\eta_z$, where $H^\pm=\kb{\phi_i^\pm|H_{\text{edge},l}|\phi^\pm_j}$ and $\phi^\pm_i$ is the eigenvectors of $\Pi$ with eigenvalue of $\pm 1$. When the eigenvalue of  $\Pi$ equal to $+1$ we obtain
\begin{equation}
    H_{\text{edge},l}^{+}=iA(l)\tilde{\eta}_{z}\partial_l+\tilde{M}^+(l)\tilde{\eta}_{x}\label{b3},
\end{equation}
where $\tilde{M}^+(l=\text{I-IV})=\{\Delta_0+J_0^z|m|/t_x, -\Delta_0-J_0^z|m|/t_y, -\Delta_0+J_0^z|m|/t_x, \Delta_0-J_0^z|m|/t_y\}$.
For the case of $- 1$ we obtain
\begin{equation}
    H_{\text{edge},l}^{-}=iA(l)\tilde{\eta}_{z}\partial_l+\tilde{M}^-(l)\tilde{\eta}_{x},
\end{equation}
where $\tilde{M}^-(l=\text{I-IV})=\{-\Delta_0+J_0^z|m|/t_x, \Delta_0-J_0^z|m|/t_y, \Delta_0+J_0^z|m|/t_x, -\Delta_0-J_0^z|m|/t_y\}$.  When $J_0^z|m|/t_0>\Delta_0$, the signs of the Dirac masses of the adjacent edges in $H_{\text{edge},l}^{\pm}$  are opposite (We first consider the isotropic case with $t_0=t_x=t_y$.) and two domain walls are formed around each corner.  Although the $x$ and $y$ components of  Ne\'{e}l vector do not produce Dirac mass on the edge states, the Dirac mass generated from the $z$ component of Ne\'{e}l vector changes as the polar angle.  The variation of Dirac masses $\tilde{M}^+(l)$ and $\tilde{M}^-(l)$ with the polar angle is shown in Figs. \ref{fig:rot} (c) and (d). The yellow region indicates the signs of the Dirac masses on the adjacent edges are opposite. Correspondingly, the phase diagram in the plane of $(\theta,\varphi)$ is shown
in Fig. \ref{fig:rot}(b). The yellow area indicates the HOTSC phase while the light blue area indicates the absence of MCMs in the system.

In the above, we have established that the in-plane component of the Ne\'{e}l vector does not generate a Dirac mass on the edge states. Now, we focus solely on exploring the effect of the $z$ component Ne\'{e}l vector with $\theta=0$. We can obtain the edge Hamiltonian from the general expression in Eq.~\eqref{pb1}  with only the $z$ component of Ne\'{e}l vector taken into account
\eqa{
    H_{\rm I}&=A_yk_y\eta_z-\Delta_0 \gamma_y\eta_y+\frac{J^z_0m}{t_x} \gamma_z\eta_x,\\
    H_{\rm II}&=-A_xk_x\eta_z+\Delta_0 \gamma_y\eta_y-\frac{J^z_0m}{t_y} \gamma_z\eta_x,\\
    H_{\rm III}&=-A_yk_y\eta_z+\Delta_0 \gamma_y\eta_y+\frac{J^z_0m}{t_x} \gamma_z\eta_x,\\
    H_{\rm IV}&=A_xk_x\eta_z-\Delta_0 \gamma_y\eta_y-\frac{J^z_0m}{t_y} \gamma_z\eta_x.\label{eq:edge2}
}
The edges spectrum read
\eqa{
    &E_{\rm I,III}(k_y)=\pm\sqrt{(A_yk_y)^2+(\mp\Delta_0\pm \frac{J_0^zm}{t_x})^2},\\
    &E_{\rm II,IV}(k_x)=\pm\sqrt{(A_xk_x)^2+(\pm\Delta_0\pm \frac{J_0^zm}{t_y})^2}.\label{eq:bryspec}
}

Taking the counterclockwise direction as positive, we can define an 'edge coordinate' $l$ as shown in Fig.~\ref{fig:ill}(a), and the effective Hamiltonian can be expressed as
\begin{equation}
    H_{\text{edge},l}=iA(l)\eta_{z}\partial_l+M(l)\gamma_{y}\eta_{y}+J(l)\gamma_z\eta_{x}\label{pb2},
\end{equation}
with the coefficients $A(l)=\{A_y,A_x,A_y,A_x\}$, $M(l)=\cl{-\Delta_0,\Delta_0,\Delta_0,-\Delta_0}$ and $J(l)=\left\{J_0^zm/t_x,-J_0^zm/t_y\right.$, $\left.J_0^zm/t_x,-J_0^zm/t_y\right\}$ for $l=\{$I, II, III, IV$\}$, respectively.
The Hamiltonian $H_{\text{edge},l}$ can be decoupled as $H_{\text{edge},l}=H_{\text{edge},l}^{+}\oplus H_{\text{edge},l}^{-}$ with the help of  $\Pi =\gamma_x\eta_z$. For $\gamma_x\eta_z=+1$ we obtain
\begin{equation}
    H_{\text{edge},l}^{+}=iA(l)\tilde{\eta}_{z}\partial_l+\tilde{M}^+(l)\tilde{\eta}_{x},\label{eq:mp}
\end{equation}
where $\tilde{M}^+(l=\text{I-IV})=\{\Delta_0+J_0^z|m|/t_x, -\Delta_0-J_0^z|m|/t_y, -\Delta_0+J_0^z|m|/t_x, \Delta_0-J_0^z|m|/t_y\}$.
For $\gamma_x\eta_z=-1$ we obtain
\begin{equation}
    H_{\text{edge},l}^{-}=iA(l)\tilde{\eta}_{z}\partial_l+\tilde{M}^-(l)\tilde{\eta}_{x}\label{eq:mm},
\end{equation}
where $\tilde{M}^-(l=\text{I-IV})=\{-\Delta_0+J_0^z|m|/t_x, \Delta_0-J_0^z|m|/t_y, \Delta_0+J_0^z|m|/t_x, -\Delta_0-J_0^z|m|/t_y\}$.
Without loss of generality, $J_0^z$, $\Delta_0$, $t_{x,y}$ and $A_{x,y}$ are taken as positive.      The signs of Dirac masses of all the adjacent edges for $H_{\text{edge},l}^{\pm}$ are opposite when $J_0^z|m|/t_x>\Delta_0$, which results in a mass domain wall and gives rise to the two MCMs at each corner~\cite{Jackiw}.  Since the edge Hamiltonian  $H^+$ and $H^-$ are decoupled from each other with the chemical potential $\mu=0$, two MZMs can still exist at each corner despite the time-reversal symmetry being broken.

%----------------------------------------------------------------------
\subsection{Isosceles triangle geometry}
In this subsection, we present the edge theory of the isosceles triangle geometry  (Fig.~\ref{fig:ill}(b)). Specifically, for edges I and IV, the effective Hamiltonian remains the same as that of the square geometry
\eqa{
    H_{\rm I}&=A_yk_y\eta_z-\Delta_0 \gamma_y\eta_y+\frac{J^z_0m}{t_x} \gamma_z\eta_x,\\
    H_{\rm IV}&=A_xk_x\eta_z-\Delta_0 \gamma_y\eta_y-\frac{J^z_0m}{t_y} \gamma_z\eta_x.
}
Concerning edge V, it should be noted that the Dirac mass term contribution stemming from the $s$-wave electron pairing is $\Delta_0$. However, in the case of altermagnets, the magnetic order spin splitting $J(\bfk)$ becomes zero in the $k_x=k_y$ direction. As a result, there is no contribution to the Dirac mass term from the altermagnets on the hypotenuse V of the isosceles right triangle~\cite{wang_high-temperature_2018}. Accordingly, we can obtain the effective Hamiltonian  on edge V as follows
\eq{
    H_{\rm V}=A_{\rm V}k_{\rm V}\eta_z-\Delta_0\gamma_y\eta_y.
}
Likewise, by utilizing $\Pi=\gamma_x\eta_z$, one can decouple the  Hamiltonian as $H=H^+\oplus H^-$. The corresponding mass terms associated with each subblock are expressed as
\eqa{
    &\tilde{M}^+(l=\text{I,V,IV})=\cl{\Delta_0+\frac{J_0^z|m|}{t_x},\Delta_0,\Delta_0-\frac{J_0^z|m|}{t_y}},\\
    &\tilde{M}^-(l=\text{I,V,IV})=\cl{-\Delta_0+\frac{J_0^z|m|}{t_x},-\Delta_0,-\Delta_0-\frac{J_0^z|m|}{t_y}}.\label{eq:trimass}
}
The interplay between electron pairing $\Delta_0$ and altermagnetism $J_0^z$ at the edge can result in the formation of  MCMs,  which will be discussed later.

    \subsection{The effective boundary Hamiltonian for the heterostructures of altermagnets with $g$-wave and $i$-wave spin splitting}
    For the heterostructures of altermagnets with the spin splitting of other symmetries, including $g$-wave, and $i$-wave~\cite{smejkal_beyond_2022-1},  the boundary Hamiltonian can be obtained by rotating the coordinate system.  Within the rotated  system,  the boundary Hamiltonian reads as
    \eq{
        H(\alpha)=A_0k_\parallel\eta_0\gamma_z+\mathcal{L}(\alpha)\eta_z\gamma_x+\Delta_0\cs{\sin \alpha \eta_y\gamma_y+\cos\alpha\eta_x\gamma_y},\label{eq:giwave}
    }
    where
    \eq{
        \left\{
        \begin{array}{c}
            \mathcal{L}(\alpha)=J_{\rm eff}\sin(4\alpha)\qquad g-\text{wave},\\
            \mathcal{L}(\alpha)=J_{\rm eff}\sin(6\alpha)\qquad i-\text{wave},
        \end{array}
        \right.
    }
    and $\gamma_i,\eta_j$ are Pauli matrices, $J_{\rm eff}$ is the effective strength of the altermagnet on the boundary. In the Eq.~\eqref{eq:giwave}, $\alpha$ is the angle of rotation of the coordinate axis, which is the direction of the tangential boundary, as shown in Fig.~\ref{fig:ill}(c). The second and the third terms are the Dirac mass terms from the altermagnet and $s$-wave electron pairing, respectively.
    The edge spectra read
    \eq{
        E(\alpha)=\pm\sqrt{A_0^2k_\parallel^2+\cs{\mathcal{L}(\alpha)\pm \Delta_0}^2}.
    }
    The two mass terms compete with each other since they are commutative. This competition leads to a topological phase transition on the boundary, which creates a mass domain wall capable of binding MCMs in the direction of $|\mathcal{L}(\alpha)|=\Delta_0$.

%=======================================================
\section{Effect of chemical potential}\label{sec:chem}
In this section, we explore the impact of finite chemical potential on MCMs. Specifically, in investigating the chemical potential $\mu\neq0$, we adopt a similar approach in that chemical potential is treated as a perturbation as the one used in Section ~\ref{sec:neel}.  We obtain the edge Hamiltonian
\begin{equation}
    H_{\text{edge},l}=iA(l)\eta_{z}\partial_l+M(l)\gamma_{y}\eta_{y}+J(l)\gamma_z\eta_{x}-\mu(l)\gamma_z,\label{eq:muedge}
\end{equation}
where the coefficients $A(l)=\{A_y,A_x,A_y,A_x\}$, $M(l)=\cl{-\Delta_0,\Delta_0,\Delta_0,-\Delta_0}$, $J(l)=\left\{J_0^z|m|/t_x,-J_0^z|m|/t_y\right.$, $\left.J_0^z|m|/t_x,-J_0^z|m|/t_y\right\}$ and $\mu(l)=\mu_0$ for $l=\{$I, II, III, IV$\}$, respectively.

For $\mu\neq0$, there is no such  $\Pi$ matrix that can decompose the edge Hamiltonian in Eq.~\eqref{eq:muedge} into $H^+\oplus H^-$. This indicates that a non-zero chemical potential brings about coupling between the two parts. Two MZMs localized at the same position would mix and vanish.
\begin{figure}
    \centering
    \includegraphics[scale=0.8]{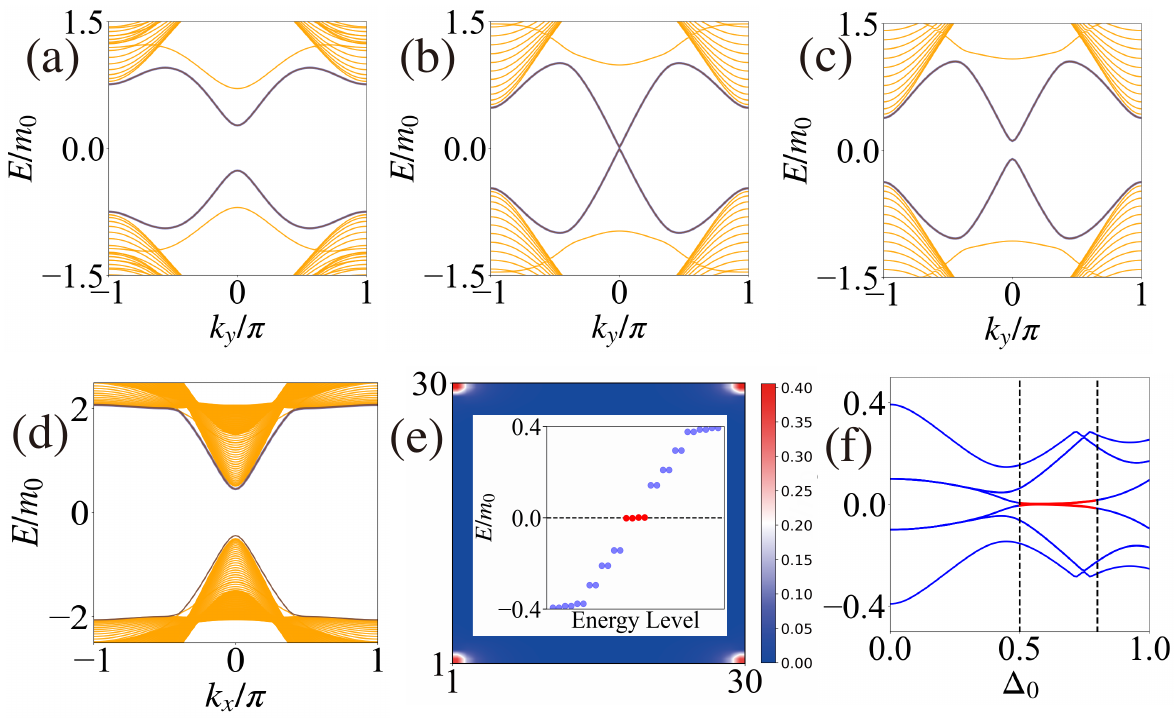}
    \caption{Quasiparticle spectrum for a cylinder geometry under different pairing strengths. For (a) $\Delta_0=0.1$. (b) $\Delta_0=0.5$. (c) $\Delta_0=0.6$. (d) Bulk spectrum and edge spectrum for the open boundary conditions along the $x$ direction with the critical electron pairing $\Delta_0=0.5$. (e) The density plot displays the corner localized probability distribution of the four MCMs.  The insert shows real space eigenvalues for $\Delta_0=0.6$. (f) Variation of real space spectrum with pairing intensity $\Delta_0$ with the MCMs labeled in red. The common parameters are: $m_0=1.0, t_y=A_y=1.0,t_x=A_x=2.0, J_0=0.5,\mu=0.1$. }\label{fig:d0}
\end{figure}

\begin{figure}
    \centering
    \includegraphics[scale=0.8]{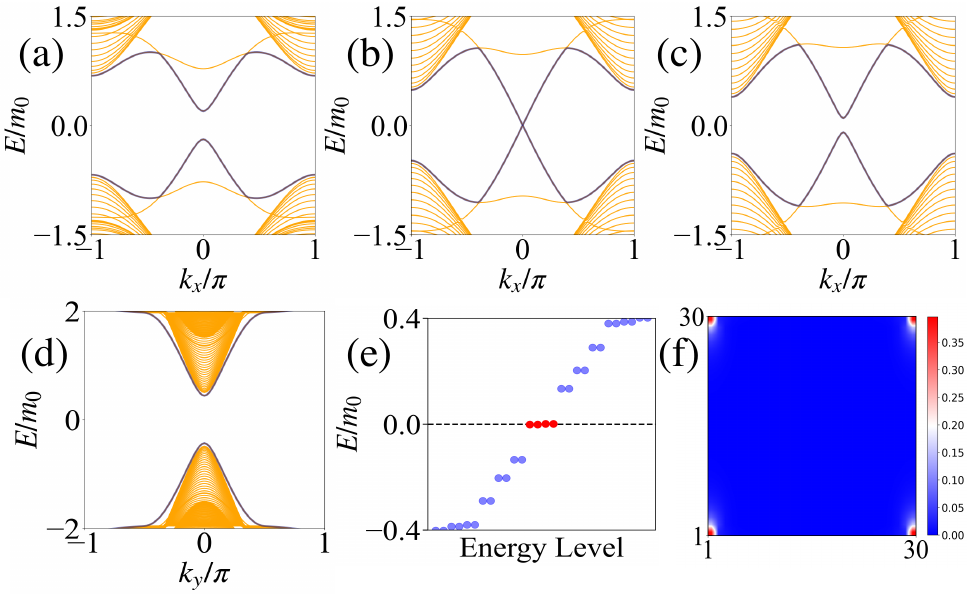}
    \caption{Quasiparticle spectrum for a cylinder geometry under different pairing strengths. For (a) $\Delta_0=0.3$. (b) $\Delta_0=0.5$. (c) $\Delta_0=0.6$. (d) Bulk spectrum and edge spectrum for the open boundary conditions along the $x$ direction with the critical electron pairing $\Delta_0=0.5$. (e)The real space eigenvalues for $\Delta_0=0.6$.  (f) The density plot displays the corner localized probability distribution of the four zero-energy MCMs.  The common parameters are: $m_0=1.0, t_y=A_y=2.0,t_x=A_x=1.0, J_0=0.5,\mu=0.1$. }\label{fig:d02}
\end{figure}

Assuming $t_x> t_y$ and selecting a finite chemical potential, we observe that as the electron pairing becomes larger, the gap of the edge states along the $k_y$ direction is closed and reopened with the gap of the bulk state and the gap of the edge states along the $k_x$ direction always open, as depicted in Figs. \ref{fig:d0}(a)-(d), which indicates a higher-order topological phase transition. In fact, according to Eq.~\eqref{eq:bryspec}, it can be concluded that once anisotropy occurs ($t_x\neq t_y$), the edge states $E(k_{x(y)})=0$ in the $x$ and $y$ directions have different critical $\Delta_0^c$. This observation is also in line with the energy spectrum of the cylinder geometry obtained via numerical calculations. A similar pattern is likewise observed in the real space spectrum for the variation of electron pairing $\Delta_0$ (as shown in Fig.~\ref{fig:d0}(f)), where one MCM per corner localization takes place after a higher-order topological phase transition, and these states subsequently are split again as the electron pairing continues to increase.

For the case where $t_x<t_y$, we observe that as the strength of electron pairing increases, a higher-order topological phase transition occurs but with the gap of the edge states along the $k_y$ direction close and reopen, as depicted in Figs. \ref{fig:d02}(a)-(d). The hallmark MZMs emerge at the corners of the system, as shown in Figs.~\ref{fig:d02}(e) and (f).

%========================================================
\section{Manipulation of MCMs}\label{sec:braiding}
\subsection{Square geometry}
In this subsection, we will first manipulate the MCMs in square geometry by using uniaxial stress. It is useful to define the corner position, corner i-j, which is defined as the intersection of the adjacent edge i and edge j. From the picture of mass terms in edge theory, we can know that the order between $J_0^z|m|/t_x$, $J_0^z|m|/t_y$, and $\Delta_0$ is important. In fact, there are six sequences, but only four are not equivalent, as shown in the following four cases. The chemical potential will couples the $H^+$ and $H^-$ subblocks. This coupling causes two MCMs from the two subblocks at the same corner to mix and annihilate. However, such two MCMs are intact if they are located at different corners.
\begin{figure}[h]
    \centering
    \includegraphics[scale=0.8]{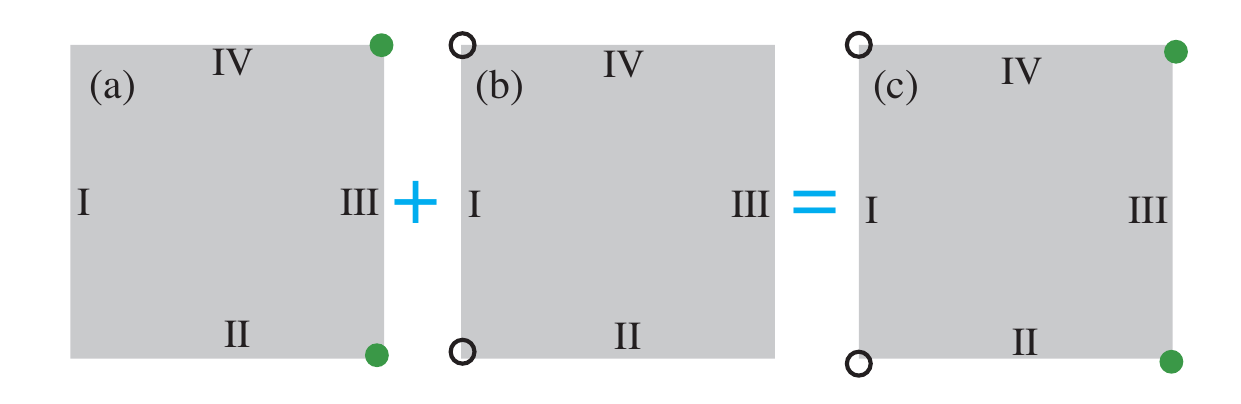}
    \caption{Schematic diagram (a) shows the $H^-$ subblock generating MCMs at corner III-IV and corner III-II. (b) shows $H^+$ subblock producing MCMs at corner I-II and corner I-IV.  (c)  The MCMs generated by $H^+$ and $H^-$ are spatially separated, so that stable MCMs can emerge after the non-zero chemical potential couples $H^+$ and $H^-$. }\label{fig:square}
\end{figure}

\begin{itemize}
    \item Case (1): $\mu\neq 0, J_0^z|m|/t_x<\Delta_0<J_0^z|m|/t_y$.
\end{itemize}

Firstly, we consider the case where $t_x>t_y$ and $\mu\neq 0$. When the electron pairing satisfies $J_0^z|m|/t_x<\Delta_0<J_0^z|m|/t_y$, one can obtain the corresponding signs of the mass terms from Eq.~\eqref{eq:mp} and Eq.~\eqref{eq:mm}
\eq{
    \text{Sign}(\tilde{M}^+)=\cl{+,-,-,-},\qquad
    \text{Sign}(\tilde{M}^-)=\cl{-,-,+,-}\label{eq:s1}.
}

The Eq.~\eqref{eq:s1} indicates that the MCMs caused by $H^+$ emerge at the corner I-IV and corner I-II, as illustrated in Fig. \ref{fig:square}(b). On the other hand, the MCMs at corner III-IV and corner III-II are due to $H^-$, as depicted in Fig. \ref{fig:square}(a). While $H^+$ and $H^-$ are coupled in the presence of a nonzero chemical potential, the final four MCMs are still stable since they are located at distinct corners. Consequently, when $J_0^z|m|/t_x<\Delta_0<J_0^z|m|/t_y$, one stable MZM is localized at each corner, as shown in Fig. \ref{fig:square}(c). This analysis is in agreement with the numerical results presented in Section~\ref{sec:chem}.
\begin{figure}[h]
    \centering
    \includegraphics[scale=0.8]{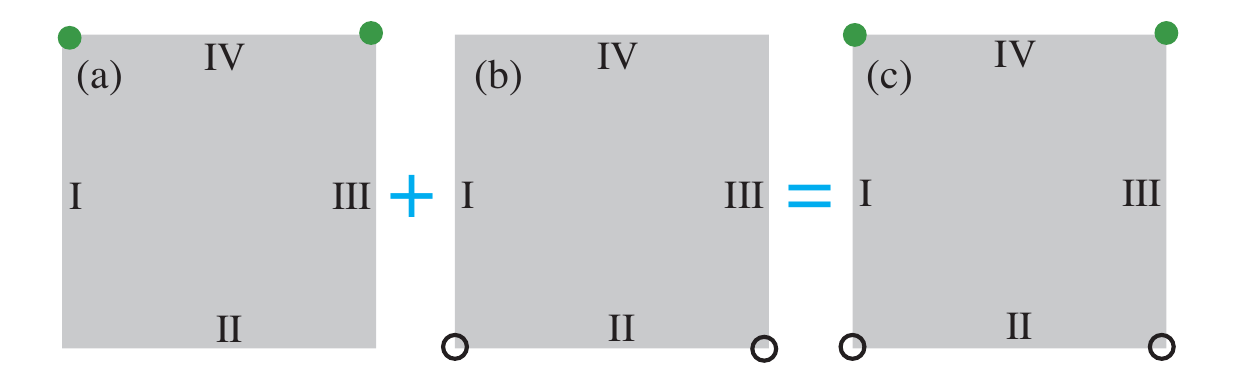}
    \caption{Schematic diagram (a) shows $H^-$ subblock generating MCMs at corner IV-I and corner IV-III. (b) shows $H^+$ subblock giving birth to MCMs at corner II-IV and corner II-III.  (c)  The MCMs generated by $H^+$ and $H^-$ are spatially separated, so that stable MCMs can emerge despite the non-zero chemical potential.}\label{fig:square2}
\end{figure}

\begin{itemize}
    \item Case (2) : $\mu\neq 0, J_0^z|m|/t_y<\Delta_0<J_0^z|m|/t_x$.
\end{itemize}

The values of $t_x$ and $t_y$ can be controlled experimentally through the application of uniaxial stress.  Now we consider $J_0|m|/t_y<\Delta_0<J_0|m|/t_x$, the signs of Dirac masses of  the edge Hamiltonian  $H^\pm$ are expressed as
\eq{
    \text{Sign}(\tilde{M}^+)=\cl{+,-,+,+},\qquad    \text{Sign}(\tilde{M}^-)=\cl{+,+,+,-}.
}
One can obtain that $H^+$ leads to the formation of MCMs at corner II-I and corner II-II, as shown in Fig.~\ref{fig:square2}(b). In contrast, the MCMs induced by the $H^-$ term appear at corner IV-I and corner IV-III, as illustrated in Fig.~\ref{fig:square2}(a). As previously noted, since the MCMs generated by the two blocks occupy distinct corners, they do not overlap and can form stable MCMs even when the two subspaces are coupled with $\mu\neq0$.
\begin{figure}[h]
    \centering
    \includegraphics[scale=0.8]{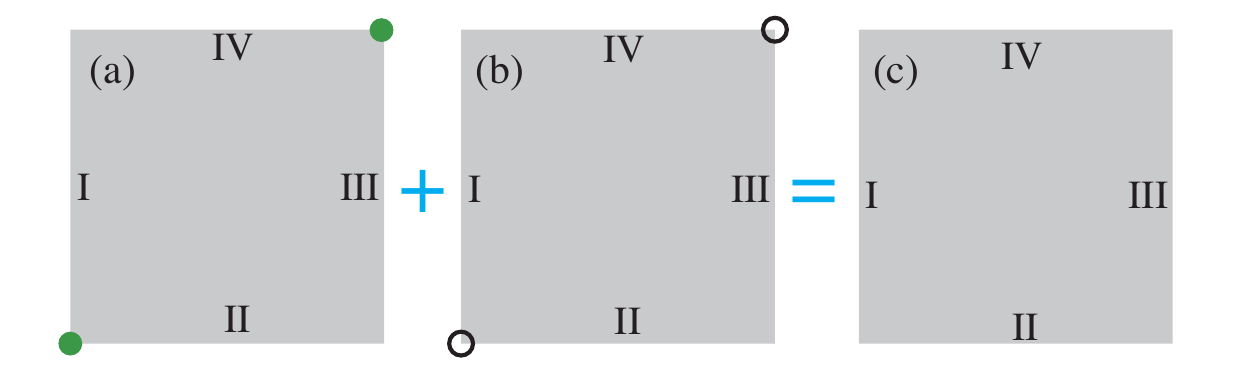}
    \caption{Schematic diagram (a) shows $H^-$ subblock generating MCMs at corner III-IV and corner I-II. (b) shows $H^+$ subblock bounding MCMs at the same corners.  (c)  Since the MCMs generated by $H^+$ and $H^-$ are at the same position, the non-zero chemical potential coupling will cause the MCMs to mix and vanish. }\label{fig:square3}
\end{figure}
\begin{itemize}
    \item Case (3) : $\mu\neq 0, J_0^z|m|/t_x\neq J_0^z|m|/t_y<\Delta_0$.
\end{itemize}

We now examine the scenario where the electron pairing satisfies $J_0|m|/t_x\neq J_0|m|/t_y<\Delta_0$. The signs of Dirac masses on the edges are expressed as
\eq{
    \text{Sign}(\tilde{M}^+)=\cl{+,-,-,+},\qquad
    \text{Sign}(\tilde{M}^-)=\cl{-,+,+,-}\label{eq:case3}.
}
Equation \eqref{eq:case3} reveals that the MCMs generated by the two subblocks $H^\pm$ emerge at the same location, as illustrated in Fig. \ref{fig:square3}. Consequently, the finite chemical potential causes $H^+$ and $H^-$ coupled, and the otherwise pair of MCMs will mix and disappear.

\begin{figure}[h]
    \centering
    \includegraphics[scale=0.8]{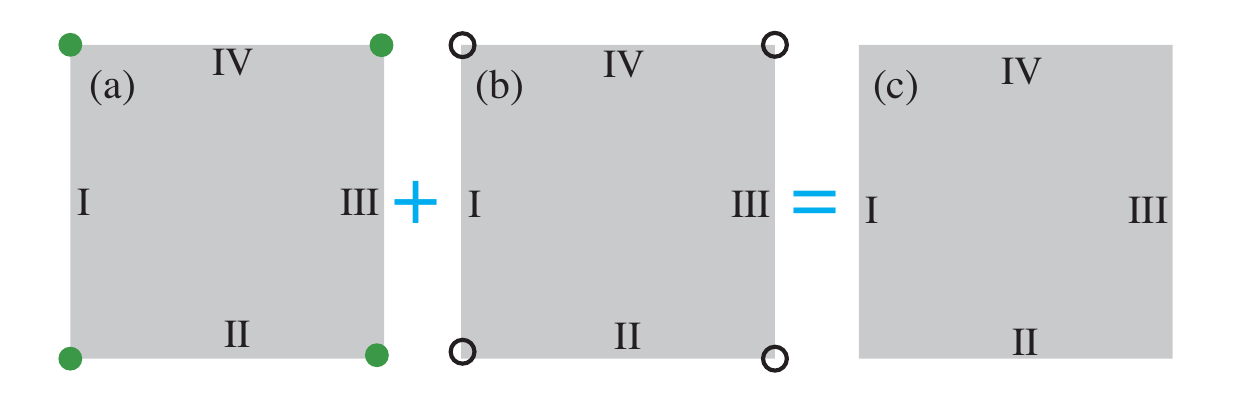}
    \caption{(a) (b) The four MCMs at four corners from $H^-$ subblock and $H^+$ subblock, respectively. Since the MCMs generated by $H^+$ and $H^-$ are at the same position, the non-zero chemical potential coupling will cause the MCMs to mix and vanish.}\label{fig:square-x}
\end{figure}
\begin{itemize}
    \item Case (4) : $\mu\neq 0, \Delta_0<J_0^z|m|/t_x\neq J_0^z|m|/t_y$.
\end{itemize}
Finally, we consider the case where $\Delta_0<J_0|m|/t_x\neq J_0|m|/t_y$. The signs of the Dirac masses term can be expressed as
\eq{
    \text{Sign}(\tilde{M}^+)=\cl{+,-,+,-},\qquad
    \text{Sign}(\tilde{M}^-)=\cl{+,-,+,-}\label{eq:case4}.
}
We find that the MCMs contributed by $H^\pm$ subblocks are located at the same corners. The non-zero chemical potential coupling $H^+$ and $H^-$ makes them mix and vanish, as shown in Fig. \ref{fig:square-x}.

\begin{figure}[h]
    \centering
    \includegraphics[scale=0.8]{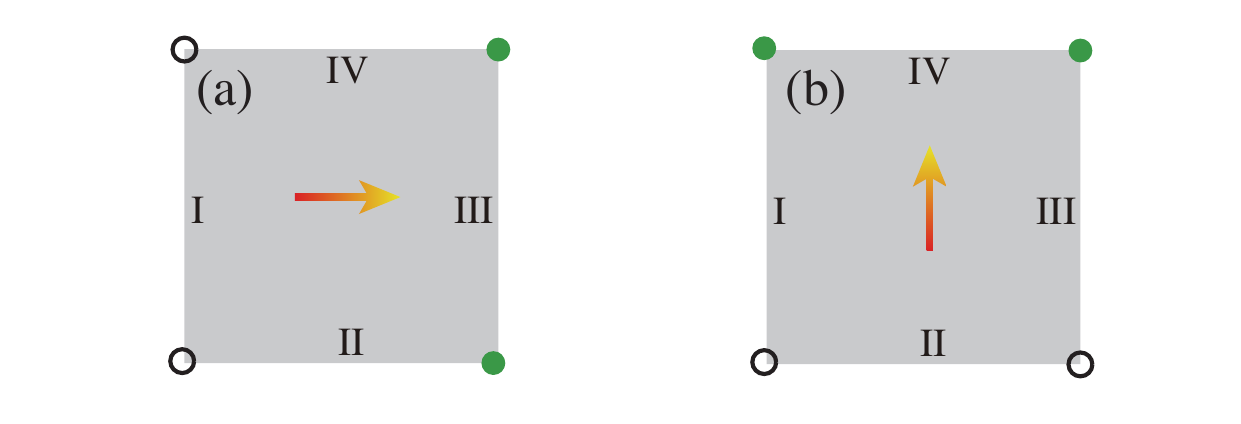}
    \caption{Tunable patterns and interchange of MCMs. (a) When $J_0^z|m|/t_x<\Delta_0<J_0^z|m|/t_y$,  two MCMs (blue hollow circles) emerge located at the left two corners from $H^+$, whereas two MCMs (green solid circles) emerge located at the right two corners from $H^-$.  (b) When $J_0^z|m|/t_x<\Delta_0<J_0^z|m|/t_y$, two MCMs (blue hollow circles) emerge located at the lower two corners from $H^+$, whereas two MCMs (green solid circles) emerge located at the upper two corners from $H^-$. The arrow indicates the uniaxial stress direction.}\label{fig:square4}
\end{figure}

Interestingly, we find that it is possible to manipulate MZMs by adjusting the relative magnitudes of $t_x$ and $t_y$ via uniaxial stress.  As shown in Figs.~ \ref{fig:square4}(a) and (b), the arrows indicate the direction of the uniaxial stress. We can interchange a pair of MCMs by applying uniaxial stresses in different directions.
%---------------------------------------------------------------------------
\subsection{Isosceles right triangle geometry}
Similar to the above subsection, we investigate how MCMs evolve in an isosceles right triangle geometry.
\begin{itemize}
    \item Case (1): $\mu\neq0, 0<\Delta_0<J_0^z|m|/t_x,\Delta_0<J_0^z|m|/t_y$.
\end{itemize}

\begin{figure}[h]
    \centering
    \includegraphics[scale=0.8]{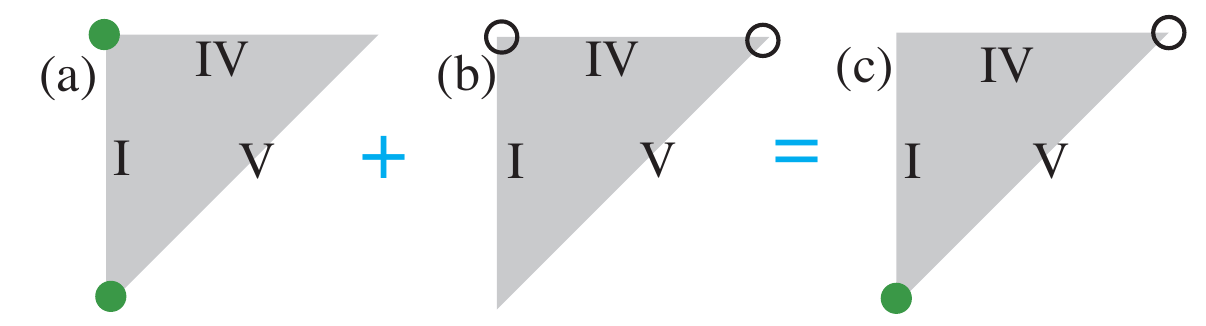}
    \caption{Schematic diagram (a) shows $H^-$ subblock generating MCMs at corner I-IV and corner I-V. (b) shows $H^+$ subblock producing MCMs at corner IV-I and corner IV-V.  (c)  The non-zero chemical potential couples $H^+$ and $H^-$, the MCMs in the same position are mixed and disappear. }\label{fig:tri1}
\end{figure}

According to Eq.~\eqref{eq:trimass}, the signs of the Dirac mass terms on the triangle edge can be expressed as follows in the case
\eq{
    \text{Sign}\cm{\tilde{M}^+(l=\text{I,V,IV})}=\cl{+,+,-},\qquad
    \text{Sign}\cm{\tilde{M}^-(l=\text{I,V,IV})}=\cl{+,-,-}\label{eq:tris1}.
}
From Eq.~\eqref{eq:tris1} we obtain that the MCMs produced by $H^+$ appear at the corner IV-V and corner I-IV, as depicted in Fig.~\ref{fig:tri1}(b). Similarly, for $H^-$, the MCMs emerge at the corner I-IV and corner I-V, as illustrated in Fig.~\ref{fig:tri1}(a).
When a finite chemical potential is present, $H^+$ and $H^-$ can couple with each other, causing the MCMs located at the same spatial location to mix and vanish, while MCMs at different locations can remain exist. Therefore, only two MCMs can exist stably, as illustrated in Fig. \ref{fig:tri1}(c), which is in agreement with the numerical results presented in the main text.
\begin{itemize}
    \item Case (2):  $\mu\neq0, 0<J_0^z|m|/t_x<J_0^z|m|/t_y<\Delta_0$.
\end{itemize}

The signs of the Dirac masses at the boundary can be expressed as follows in the case
\eq{
    \text{Sign}\cm{\tilde{M}^+(l=\text{I,V,IV})}=\cl{+,+,+},\qquad
    \text{Sign}\cm{\tilde{M}^-(l=\text{I,V,IV})}=\cl{-,-,-}\label{eq:tris2}.
}
As can be obtained from Eq.~\eqref{eq:tris2}, there is no domain wall, which means that MCMs cannot emerge.

\begin{figure}[h]
    \centering
    \includegraphics[scale=0.8]{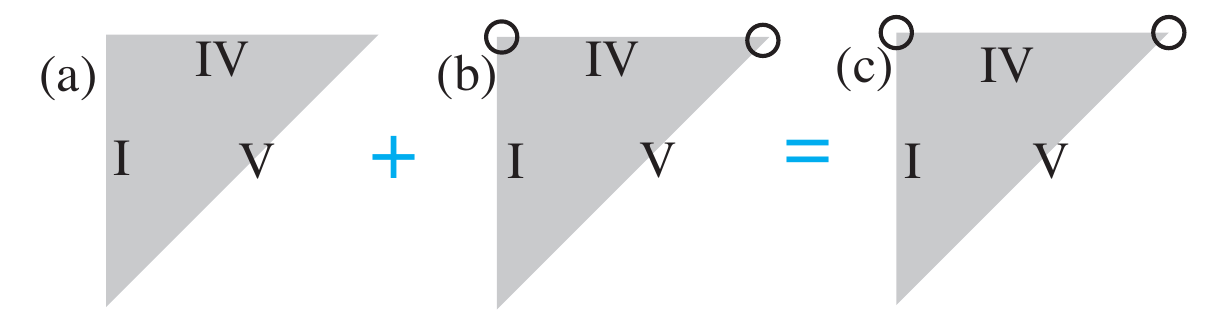}
    \caption{Schematic (a) indicates that the $H^-$ subblock does not contribute to the MCMs. (b) shows $H^+$ subblock giving birth to MCMs at corner IV-I and corner IV-V. (c)  The non-zero chemical potential couples $H^+$ and $H^-$.}\label{fig:tri2}
\end{figure}
\begin{figure}[h]
    \centering
    \includegraphics[scale=0.8]{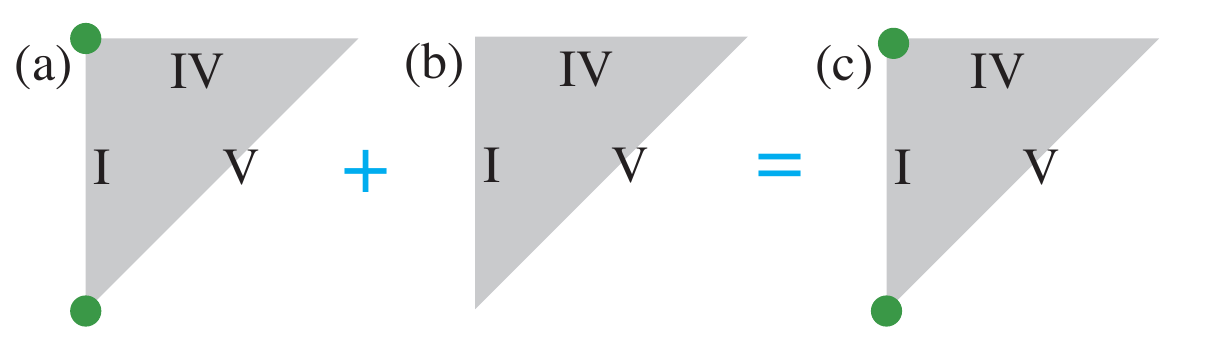}
    \caption{ Schematic (a) indicates that the  $H^-$ subblock generating MCMs at corner V-IV and corner V-I. (b) shows that the $H^+$ subblock does not generate MCMs.  (c)  The non-zero chemical potential couples $H^+$ and $H^-$.}\label{fig:tri3}
\end{figure}
\begin{itemize}
    \item Case (3): $\mu\neq0, J_0|m|/t_x<\Delta_0<J_0|m|/t_y$.
\end{itemize}
The signs of the Dirac masses on the edges in the case are expressed as
\eq{
    \text{Sign}\cm{\tilde{M}^+(l=\text{I,V,IV})}=\cl{+,+,-},\qquad
    \text{Sign}\cm{\tilde{M}^-(l=\text{I,V,IV})}=\cl{-,-,-}\label{eq:tris3}.
}
Upon analyzing the signs of the mass terms, one can observe that all domain walls are generated by the $H^+$ subblock. The MCMs emerge and are stable at corner IV-I and corner IV-V, as illustrated in Fig.~\ref{fig:tri2}.

\begin{itemize}
    \item Case (4): $\mu\neq0, J_0|m|/t_y<\Delta_0<J_0|m|/t_x$.
\end{itemize}

In this scenario, the signs of the Dirac masses can be expressed as follows:
\eq{
    \text{Sign}\cm{\tilde{M}^+(l=\text{I,V,IV})}=\cl{+,+,+},\qquad
    \text{Sign}\cm{\tilde{M}^-(l=\text{I,V,IV})}=\cl{+,-,-}\label{eq:tris4}.
}
According to Eq.~\eqref{eq:tris4}, the MCMs are generated solely by $H^-$, as illustrated in Fig.~\ref{fig:tri3}. Hence,  non-zero chemical potential does not work with two stable MCMs finally.

In summary, the movement of a pair of MCMs in space is achieved by uniaxial stress in the isosceles triangle structure.


\begin{thebibliography}{89}
\bibitem [{sup({\natexlab{b}})}]{supp2}%
    \BibitemOpen
    \href@noop {} {\bibfield  {journal} {\bibinfo  {journal} {The in-plane
                N\'{e}el vector does not generate a Dirac mass on the edge.}\ }
        ({\natexlab{b}})}\BibitemShut {NoStop}%
\end{thebibliography}
\end{document}